\documentclass[12pt]{article}
\usepackage{amsmath,amssymb,amsfonts,epsfig,graphicx,euscript}%
\usepackage[bookmarks,bookmarksnumbered,linktocpage,pdfstartview=FitH]{hyperref}
\hypersetup{colorlinks,%
citecolor=red,%
filecolor=blue,%
linkcolor=blue,%
urlcolor=blue,%
pdftex}
\usepackage{amsmath}% http://ctan.org/pkg/amsmath
\usepackage{amsfonts}% http://ctan.org/pkg/amsfonts
\usepackage{graphicx}
\usepackage{caption}
\usepackage{subcaption}
\usepackage{float}
\usepackage[all]{hypcap}
\numberwithin{equation}{section}
\usepackage{cite}
\usepackage{calc}
\newlength{\spacer}
\newsavebox{\mybox}

\newcommand{\bse}{\begin{subequations}}
\newcommand{\ese}{\end{subequations}}
\newcommand{\be}{\begin{equation}}
\newcommand{\ee}{\end{equation}}
\newcommand{\bea}{\begin{eqnarray}}
\newcommand{\eea}{\end{eqnarray}}
\newcommand{\ba}{\begin{array}}
\newcommand{\ea}{\end{array}}

\renewcommand{\thefootnote}{\fnsymbol{footnote}}

\begin{document}
%\hfill%
%\vbox{
%    \halign{#\hfil        \cr
%           IPM/P-2012/010\cr
%                     }
%      }
%\vspace{1cm}
\begin{center}
%{ \large{\textbf{The Baryonic Contribution to the U$_\textrm{Y}$(1) Chern-Simons Term and its effect on the Evolution of Fermionic Asymmetries and Hypermagnetic Fields  }}} %\\
{ \large{\textbf{
%The Effect of Baryonic Contribution to the U$_\textrm{Y}$(1) Chern-Simons Term on the Evolution of Fermionic Asymmetries and Hypermagnetic Fields
The effects of the U$_\textrm{Y}$(1) Chern-Simons term and its baryonic contribution on matter asymmetries and hypermagnetic fields  
}}} %\\
\vspace*{1.5cm}
\begin{center}
{\bf  S. Rostam Zadeh\footnote{S$_{-}$Rostamzadeh@sbu.ac.ir}, S. S. Gousheh\footnote{ss$-$gousheh@sbu.ac.ir}}\\%
\vspace*{0.4cm}
{\it {Department of Physics, Shahid Beheshti University, G.C., Evin, Tehran 19839, Iran}}  \\
\vspace*{1cm}
\end{center}
\end{center}
\begin{center}
\today
\end{center}
%\vspace{.5cm}
%\bigskip

\renewcommand*{\thefootnote}{\arabic{footnote}}
\setcounter{footnote}{0}

\begin{center}
\textbf{Abstract}
\end{center}
In this paper, we study the significance of the U$_\textrm{Y}$(1) Chern-Simons term in general, and its baryonic contribution in particular, for the evolution of the matter asymmetries and the hypermagnetic field in the temperature range $100$GeV$\leq T \leq 10$TeV. We show that an initial helical hypermagnetic field, denoted by $B_Y^{(0)}$, can grow matter asymmetries from zero initial value. However, the growth which is initially quadratic with respect to $B_Y^{(0)}$, saturates for values larger than a critical value. The inclusion of the baryonic contribution reduces this critical value, leading to smaller final matter asymmetries. Meanwhile, $B_Y(T_{EW})$ becomes slightly larger than $B_Y^{(0)}$. In the absence of the U$_\textrm{Y}$(1) Chern-Simons term, the final values of matter asymmetries grow without saturation. Conversely, we show that an initial matter asymmetry can grow an initial seed of hypermagnetic field, provided the Chern-Simons term is taken into account. The growth process saturates when the matter asymmetry drops abruptly. When the baryonic contribution is included, the saturation occurs at an earlier time, and $B_Y (T_{EW})$ becomes larger. We also show that the baryonic asymmetry and the magnetic field strength can be within the acceptable range of present day data, provided the inverse cascade process is also taken into account; however, the magnetic field scale obtained from this simple model is much lower than the ones usually assumed for gamma ray propagation.

\newpage

\tableofcontents

\section{Introduction}\label{Introduction}
The origin of matter is still one of the great mysteries of nature. There is observational evidence that the matter in the present day Universe is the remnant of a small matter-antimatter asymmetry $\eta_B\sim 10^{-10}$ in the early Universe, i.e. just before the primordial plasma entered the hadronization phase. The value of this asymmetry has been determined independently in two different ways: first from the abundances of light elements in the intergalactic medium \cite{Fields}, and second from the power spectrum of temperature fluctuations in the Cosmic Microwave Background (CMB) \cite{Simha}.
%There is observational evidence that the present day universe has no significant amounts of baryonic antimatter, and the baryons are the remnant of a small matter-antimatter asymmetry $\eta\sim 10^{-10}$ in the early universe.\\
%There is observational evidence for a matter-antimatter asymmetry in the early universe, which leads to the remnant matter density we observe today.\\
%baryogenesis, the dynamical generation of a matter-antimatter asymmetry\\
%It was believed that antimatter is an exact mirror of matter due to the C,P and T symmetries in nature.
%Antimatter was believed to be an exact mirror of matter when it was discovered. At that time, the symmetries of parity (P), charge conjugation (C) and time reversal (T) were respected by all observed phenomena, and very little was known about the early universe. Therefore, the complete absence of antimatter (except in cosmic rays) was explained by assuming that the universe was set up like this.
%At the time of its discovery, antimatter was thought to be an exact mirror of matter; all phenomena that had been observed in nature were invariant under conjugation of parity (P) and charge (C) as well as time reversal (T), and not much was known about the early history of the universe. Henceforth, the enormous matter-antimatter asymmetry of the nearby universe (complete absence of antimatter except in cosmic rays) posed a mystery that could only be explained by assuming that the universe was set up like this.
The observational discovery of the cosmic expansion \cite{Hubble} and CMB \cite{Penzias} strengthens the big bang theory which asserts that the Universe was hot during its early stages \cite{Gamow}, and antimatter was present when pair creation and annihilation processes were in thermal equilibrium. As the temperature decreased in the plasma of the early Universe, 
%the particle energies lowered and the pair creation became rare. This led to the annihilation of 
almost all of the particles and antiparticles were annihilated and a small amount of matter remained.
%With rise of the big bang theory after the theoretical prediction [4, 5] and observational discovery of the cosmic expansion [6] and microwave background (CMB) [7], it came clear that the universe was hot during the early stages of its history [8], and antimatter was present when pair creation and annihilation reactions were in thermal equilibrium. When particle energies in the cooling plasma became too small for pair creation to occur, almost all particles and antiparticles annihilated each other, with a small amount of matter (by definition) surviving.
The discovery of C, P \cite{Wu} and CP \cite{Christenson} violation raised the possibility that the matter-antimatter asymmetry may have been created dynamically by baryogenesis, as well as leptogenesis, from an initial state which is matter-antimatter symmetric. In a seminal paper, Sakharov stated three necessary conditions for successful baryogenesis which are: the existence of baryon number violation processes, C and CP violation, and deviation from thermal equilibrium \cite{Sakharov}. The idea of baryogenesis was elevated by the paradigm of cosmic inflation \cite{Starobinsky} which states that the Universe had an accelerated expansion in its very early history explaining its spatial flatness and the isotropy of the CMB temperature. Therefore, any preexisting baryon asymmetry was diluted and negligible at the end of inflation \cite{Canetti}. 
%The discovery of violations of P [9] and CP [10] invariance (and thus also C invariance) in nature provided hints that this asymmetry may have been created dynamically by baryogenesis from a matter-antimatter symmetric initial state. The paradigm of cosmic inﬂation [13] elevated the idea of an initial state with B = 0 from an assumption, based on aesthetic reasoning, to a generic prediction. If the universe underwent a period of accelerated expansion during its very early history that lasted for long enough to explain its spacial ﬂatness and the isotropy of the CMB temperature, any pre-existing baryon asymmetry was diluted and negligible at the end of inﬂation 3. 
%Therefore, baryogenesis needs to occur either during reheating or in the radiation dominated epoch.  

A seemingly unrelated but important discovery, which can be rightfully called another great mystery of nature, was the detection of a long range magnetic field coherent over scales of the order of $30$ Kpc with a strength of order $\mu$G over the plain of the disc of the Milky Way galaxy \cite{Kronberg}. Interestingly, similar magnetic fields have been observed in other spiral and barred galaxies \cite{Kronberg2008,Bernet,Wolfe} as well as galaxy clusters \cite{Clarke, Bonafede, Feretti} and high redshift protogalactic structures \cite{Kandus}. It is generally believed that these magnetic fields are produced from the amplification of some seed fields \cite{Harrison} whose strength and origin are largely unknown \cite{Kronberg, Kulsrud}. %There are two groups of 
%models presented
%scenarios presented for the origin of these seed fields. In the astrophysical scenario, the generation of the seed fields accompanies the gravitational collapse leading to structure formation \cite{Durrer}. However, in the cosmological scenario, the seed fields are produced in the early universe during epochs before the structure formation \cite{Quashnock, Cheng, Kibble, Vachaspati, Enqvist, Baym, Grasso}. 
The fact that the magnetic fields are present ubiquitously at high redshifts, strengthens the idea that their origin is cosmological, and magnetic fields may have pervaded the Universe in its hot early stages \cite{Kandus}. The presence of coherent magnetic fields in the low density intergalactic medium, which has been reported recently \cite{Vovk, Neronov,Tavecchio,Tavecchio2011,Ando,Essey}, supports the idea of primordial magnetism as well.

%Assuming that the seed fields are primordial%(before the time of recombination)
%, they should have been generated out of thermal equilibrium 
%to be able to break the isotropy 
%\cite{Davidson}. Therefore, most of the scenarios presented for the generation of the seed fields in the early Universe operate either at a phase transition \cite{Quashnock,Cheng,Kibble,Vachaspati,Enqvist,Baym,Copi} or during the inflation \cite{Turner,Bamba,Anber,Garretson,Durrer2,Adshead,Fujita2}. The inflationary mechanisms have received a lot of attention, since they have the advantage of achieving super-horizon correlations and therefore generate much more coherent magnetic fields in the early Universe. However, due to the conformal invariance of the electromagnetism which leads to the conservation of the magnetic flux \cite{Turner}, the strength of generated magnetic fields decreases exponentially due to rapid expansion of the inflationary Universe. Different mechanisms of this kind try to solve this problem by breaking the conformal invariance in various ways \cite{Dimopoulos}. In most of these scenarios, the generated magnetic fields are helical as well (e.g. from axion dynamics during inflation). The helical magnetic fields further evolve experiencing the inverse cascade process which increases their scale in the radiation-dominated era after inflation. In this work, we assume that the helical magnetic fields are present in the symmetric phase due to an inflationary mechanism.

Assuming that the seed fields are primordial%(before the time of recombination)
, they should have been generated out of thermal equilibrium 
%to be able to break the isotropy 
\cite{Davidson}. Therefore, most of the scenarios presented for the generation of the seed fields in the early Universe operate either at a phase transition \cite{Quashnock,Cheng,Kibble,Vachaspati,Enqvist,Baym,Copi} or during the inflation \cite{Turner,Bamba,Anber,Garretson,Durrer2,Adshead,Fujita2}. The inflationary mechanisms have received a lot of attention, since they have the advantage of achieving super-horizon correlations and therefore generate much more coherent magnetic fields in the early Universe. However, the conformal invariance of the electromagnetism leads to the conservation of the magnetic flux \cite{Turner}, and hence the strength of generated magnetic fields decreases exponentially due to rapid expansion of the inflationary Universe. Various mechanisms have been proposed to solve this problem by breaking the conformal invariance \cite{Dimopoulos}. In most of these scenarios, the generated magnetic fields are helical as well (e.g. from axion dynamics during inflation). The helical magnetic fields further evolve experiencing the inverse cascade process which increases their scale in the radiation-dominated era after inflation. In this work, we assume that the helical magnetic fields are present in the symmetric phase.

%The universe in its hot early stages may have contained some magnetic fields. In fact, some  large scale magnetic fields coherent over scales of the order of $30$ Kpc have been in our galaxy \cite{Kronberg}. The strength of these fields in the Milky Way and several spiral galaxies is of the order of the microgauss \cite{Kronberg}. It is believed that some primordial seed fields whose nature is largely unknown \cite{Kronberg, Kulsrud}, are needed for the generation of galactic magnetic fields \cite{Harrison}. Seed fields might be produced during the epoch of galaxy formation, or ejected by first supernova or active galactic nuclei \cite{Semikoz, Dvornikov}. Alternatively, they might arise from phase transitions in the very early universe \cite{Quashnock, Cheng, Kibble, Vachaspati, Enqvist, Baym} down to the cosmological inflation epoch \cite{Grasso}. New signatures for the presence of cosmological magnetic fields (CMF) in the intergalactic medium, which may have survived till today, have been observed recently \cite{Vovk, Neronov}. However, weather the primordial magnetic fields have served as seed fields for the generation of the galactic magnetic fields \cite{Widrow,Beck} or long range magnetic fields observed in voids \cite{Tavecchio,Dolag}, is still an open question in cosmology.

It is well known that at high temperatures, non-Abelian long range magnetic fields cannot exist because their corresponding gauge bosons obtain a magnetic mass gap $\sim$ $g^2T$ \cite{Gross}. Thus, the only long range magnetic field surviving in the plasma is associated with the Abelian U(1) group whose vector particle remains massless \cite{Kajantie}. Moreover, electric fields decay quickly due to the large conductivity of the plasma. 
%In the symmetric phase, the long range hypercharge magnetic fields couple to the fermions chirally. 
%It is well known that at high temperatures, the non-Abelian gauge bosons contrary to the Abelian one, obtain a \textit{magnetic} mass gap $\sim$ $g^2T$. Thus, the only long range field surviving in the plasma will be the Abelian U(1) gauge boson. Moreover, electric fields quickly decay due to the finite conductivity of the plasma. The long range hypercharge magnetic fields of the symmetric phase couple to the fermions chirally.   
%During the electroweak phase transition, the long range hypercharge magnetic fields of the symmetric phase are converted to the Maxwellian magnetic fields of the broken phase\cite{Giovannini}. The ordinary electromagnetic fields couple to the fermions vector-like while, the hypercharge fields couple to the fermions chirally.
In the symmetric phase, the hypercharge fields couple to the fermions chirally. 
This leads to the fermion number violation through the Abelian anomaly, $\partial_\mu j^\mu \sim \frac{g'^2}{4\pi^2} \textbf{E}_\textbf{Y}.\textbf{B}_\textbf{Y}$. 
Here, $g'$ is the U$_\textrm{Y}$(1) gauge coupling \cite{Giovannini}. 
The anomalous coupling of the hypercharge fields to fermion number densities  appearing in the above equation, also shows up as the U$_\textrm{Y}$(1) Chern-Simons term.
%, can also be expressed through the hypermagnetic Chern-Simons term appearing in the the effective Lagrangian density of U$_\textrm{Y}$(1) gauge fields.  
 
%Hypermagnetic fields interact with matter differently from what ordinary magnetic fields do. The hypercharge fields couple to the Fermions chirally, while the coupling of the ordinary electromagnetic fields is vector-like. Consequently, the simultaneous presence of hyperelectric and hypermagnetic fields leads to the Fermion number violation due to the Abelian anomaly, $\partial_\mu j^\mu \sim \frac{g'^2}{4\pi^2} \textbf{E}_\textbf{Y}.\textbf{B}_\textbf{Y}$. Here, $g'= \frac{e}{cos\theta_W}$ is the $U_Y(1)$ gauge coupling \cite{Giovannini}. The Abelian anomaly states that the hypercharge fields are coupled to the Fermionic number density. This effect also appears as the Chern-Simons term for the corresponding gauge fields.

%It was shown long ago that, at high temperatures and finite fermion densities, the effective Lagrangian density of $SU(2)$ gauge fields acquires a Chern-Simons term \cite{Redlich}.
%It was shown long ago that, at high temperatures and finite fermion densities, a Chern-Simons term emerges in the effective Lagrangian density of SU(2) gauge fields \cite{Redlich}. Since, the Abelian hypercharge fields couple to the fermions chirally, a Chern-Simons term is induced for them as well \cite{Giovannini, Laine}.

At high temperatures and finite fermion densities, the Chern-Simons terms emerge in the effective Lagrangian densities of SU(2)$_\textrm{L}$ and U$_\textrm{Y}$(1) gauge fields due to their chiral couplings to fermions \cite{Redlich,Giovannini, Laine}. 
%The hypercharge fields and fermion number densities couple to each other  anomalously in the hypermagnetic Chern-Simons term.
%There exists the anomalous coupling between the hypercharge fields and fermion number densities in the U$_\textrm{Y}$(1) Chern-Simons term which leads to some important effects.
%The anomalous coupling between the hypercharge fields and fermion number densities also shows up in the U$_\textrm{Y}$(1) Chern-Simons term and leads to some important effects.
%The U$_\textrm{Y}$(1) Chern-Simons term leads to some important effects.
The U$_\textrm{Y}$(1) Chern-Simons term leads to the appearance of  
%The hypermagnetic Chern-Simons term clarifies that the hypercharge fields and fermion number densities couple to each other. This anomalous coupling has two important consequences. 
%First, 
a new anomalous term in the magnetohydrodynamic equations which are subsequently called the anomalous MHD (AMHD) equations \cite{Giovannini, Joyce}. 
As mentioned earlier, the evolution equations of the anomalous charge densities acquire a hypermagnetic source term as well (the Abelian anomaly). 
%The anomalous coupling between fermions and hypercharge magnetic fields \cite{Giovannini,Laine,Joyce,Rubakov} might have major effects in cosmology.
The mutual effects of the fermions and hypermagnetic fields on each other might have major effects in cosmology \cite{Giovannini,Laine,Joyce,Rubakov}. 
As a matter of fact, some authors believe that the evolutions of matter-antimatter asymmetries and the hypermagnetic field are intertwined \cite{Giovannini, Laine, Joyce,Dvornikov2011,Dvornikov,Smirnov,Sokoloff,Long,Fujita,Kamada,Long2,Kamada2,Cado}.

There exist $n_G$ global charges, i.e. $N_i = B/n_G -L_i$, which are exactly conserved in the Standard Model. Here, $n_G$ is the number of generations, $B$ is the baryon number, and $L_i$ is the lepton number of the $i$-th generation. Assigning $n_G$ chemical potentials ${\mu}_i$, $i=1,..., n_G$ to these charges, and also introducing $\mu_Y$ corresponding to the weak hypercharge which will be fixed due to the hypercharge neutrality of the plasma, $\langle Y \rangle = 0$, one can describe the electroweak plasma in complete thermal equilibrium \cite{Gorbunov}.

%The electroweak plasma in complete thermal equilibrium can be described by $n_G$ chemical potentials ${\mu}_i$, $i=1,..., n_G$ corresponding to $n_G$ exactly conserved global charges $N_i = B/n_G -L_i$. Here, $B$ is the baryon number, $L_i$ is the lepton number of the $i$-th generation and $n_G$ is the number of generations. Moreover, another chemical potential $\mu_Y$ corresponding to weak hypercharge is introduced which will be fixed as a consequence of the hypercharge neutrality of the plasma, $\langle Y \rangle = 0$ \cite{Gorbunov}.

It was discussed years ago that right-handed electrons which have a very small Yukawa coupling with Higgs bosons $h_e = 2.94 \times {10}^{-6}$ and do not take part in any weak interaction, are decoupled from the thermal ensemble at temperatures above $T_{RL} \sim 10$ TeV \cite{Campbell}. This is due to the fact that, in this range of temperatures, the rates ${\Gamma}_{RL} \sim h_e^2T$ of the relevant reactions\footnote{It is discussed in the third paper of Ref.\ \cite{Campbell} that some gauge and fermion scattering processes (such as $e_R H\leftrightarrow L_e A$, where $A=Y$ or $W$, and $e_R L_f\leftrightarrow L_e f_R$) also contribute to the chirality flip rate of electrons. 
%There are some gauge and fermion scattering processes (such as $e_R H\leftrightarrow L_e A$, where $A=Y$ or $W$, and $e_R L_f\leftrightarrow L_e f_R$) in addition to the direct and inverse Higgs decays, which participate in the chirality flip of electrons and have contributions to the total chirality flip rate (see the third paper of \cite{Campbell}).
} (direct and inverse Higgs decays in processes $e_L \bar{e}_R\leftrightarrow\phi^{(0)}$ and $\nu_e^L \bar{e}_R\leftrightarrow\phi^{(+)}$ and their conjugate processes) are much lower than the Hubble expansion rate $H \sim T^2$. 
%Thus, the electroweak theory obtains a new partially conserved charge and its related chemical potential is added to those mentioned $n_G + 1 = 4$ (for 3 generations) chemical potentials \cite{Giovannini}. 
%Thus, right-handed electron number is partially conserved if Abelian anomaly is not accounted for, and its related chemical potential is added to those mentioned $n_G + 1 = 4$ (for 3 generations) chemical potentials \cite{Giovannini}.
Thus, neglecting the Abelian anomaly, the right-handed electron number is partially conserved and its associated chemical potential can be added to the aforementioned $n_G + 1 = 4$ (for 3 generations) chemical potentials of the electroweak theory \cite{Giovannini}.
%Thus, neglecting the Abelian anomaly, the electroweak theory obtains a new partially conserved charge and the right-handed electron chemical potential is added to those mentioned $n_G + 1 = 4$ (for 3 generations) chemical potentials \cite{Giovannini}.

Considering the above fact, the authors of \cite{Campbell} suggested the following scenario in which, a right-handed electron asymmetry might preserve a primordial baryon asymmetry from the weak sphalerons: At temperatures above $T_{RL}$,\footnote{In the first paper of Ref.\ \cite{Campbell}, the value of $T_{RL}$ was computed as $T_{RL} \simeq 1$TeV.
%The value of $T_{RL}$ computed in the first paper of Ref.\ \cite{Campbell} was $T_{RL} \simeq 1$TeV.
} the weak sphalerons could not wash out the asymmetry of right handed electrons, and therefore that of baryons. However, at temperatures below $T_{RL}$, the chirality flip processes turn the right-handed electrons into left-handed leptons, while roughly at these temperatures, the weak sphalerons gradually start to fall out of equilibrium.\footnote{More accurate computations for the temperature at which the weak sphalerons fall out of equilibrium has been done recently \cite{Burnier}. 
%In recent years, the temperature at which the weak sphalerons go out of thermal equilibrium is computed more accurately \cite{Burnier}.
} Thus, it was conjectured that they might not be able to transform the left-handed leptons into antiquarks to wipe out the remaining baryon and lepton asymmetry \cite{Campbell}.
%In a scenario suggested by the authors of \cite{Campbell}, a primordial baryon asymmetry is preserved by an asymmetry in the number of right-handed electrons which are protected from washing out by sphalerons down to $T_{RL}$. Since, the sphalerons start to fall out of thermal equilibrium at temperatures close to $T_{RL}$ while the chirality flip processes start around $T_{RL}$, it is probable that the transformation of right-handed electrons into left-handed leptons is insignificant during the overlap time of these two processes. Thus, the sphalerons may not be able to turn them into antiquarks and thereby wash out the remaining baryon and lepton asymmetry \cite{Campbell}.

Afterwards in related works, the authors of \cite{Giovannini,Dvornikov2011,Dvornikov,Smirnov,Sokoloff} assumed the presence of the large scale hypermagnetic fields in the plasma, and considered the Abelian anomalous effects for right-handed electrons, which led to the generation of baryon and lepton asymmetries. The reverse effect has been studied by assuming an asymmetry for right-handed electrons while considering the Abelian anomalous effects. This situation gives rise to the 
%emergence of a hypermagnetic Chern-Simons term, and 
generation of long range hypermagnetic fields, when the full range of frequency spectrum for the hypermagnetic field is taken into account \cite{Joyce}.

In our previous work \cite{shiva}, we studied the simultaneous evolution of baryon and the first generation lepton asymmetries, and long range hypermagnetic fields, considering the Abelian anomalous effects. For that purpose, we presented the general form of the U$_\textrm{Y}$(1) Chern-Simons term which showed how chemical potentials of various fermion species contribute to it with different coefficients (see Eq.\ (2.7) of Ref.\ \cite{shiva}). Most importantly, we emphasized that the chemical potentials of right-handed and left-handed particles contribute with opposite signs to the coefficient of the U$_\textrm{Y}$(1) Chern-Simons term, in contrast to what has been used in some of the previous works. In order to explore the consequences of this one correction, we used a simple model presented in one of these works as a testing ground and implemented our correction, while keeping all other main assumptions of the model unaltered so that the results would be comparable. We then compared our results with theirs. The simplifying assumptions implemented in the model were the following: Only the contribution of the first generation leptonic chemical potentials to the U$_\textrm{Y}$(1) Chern-Simons term were considered and that of the baryonic ones was ignored. Only the electron chirality flip processes via inverse Higgs decays were considered.\footnote{None of the chirality flip reactions mentioned in footnote 1 were considered. Indeed, the inverse Higgs decays were fast enough for our investigations.} These processes violate chiral electron numbers and tend to reduce the electron chiral asymmetry,\footnote{The evolution of electron chiral asymmetry $\Delta\mu=\mu_{e_R}-\mu_{e_L}$ and Maxwellian magnetic fields are strongly coupled in the broken phase \cite{Boyarsky}, therefore the value of this asymmetry before EWPT is important.} especially when they enter into thermal equilibrium below $T_{RL}$.
Moreover, the Higgs asymmetry was assumed to be zero\footnote{This assumption leads to the absence of any net contribution from direct Higgs decays to chirality flip processes.} and also the weak sphaleron processes were neglected.\footnote{For some of the issues concerning the weak sphalerons and their consequences see Section \ref{Static Chern-Simons Terms}.} 

%A simplified form of the Chern-Simons term is used in Ref.\ \cite{Dvornikov2011} which only considers the contribution of the chemical potentials of first generation leptons and ignores that of the baryons. We presented one correction to this simplified form by using the general form of the U$_\textrm{Y}$(1) Chern-Simons term, and investigated its effect on the evolutions by comparing our results with those of that reference.
%However, we used a simplified form of this Chern-Simons term by presenting one correction to the form used in Ref.\ \cite{Dvornikov2011} which only takes into account the contribution of the chemical potentials of first generation leptons and ignores that of the baryons.
%, and then investigated its effects on the evolutions by comparing our results with those of that reference. Indeed, this simplified form only takes into account the contribution of the chemical potentials of first generation leptons and ignores that of the baryons.

As mentioned earlier, the evolution of matter asymmetries and hypermagnetic fields are strongly coupled, since they have mutual effects on one another through the Abelian anomaly and the U$_\textrm{Y}$(1) Chern-Simons term. However, in some of the previous works, the Chern-Simons term is neglected and it is assumed to be a negligible backreaction process with unimportant effects on baryogenesis and magnetogenesis. Moreover, some other former studies which have considered this Chern-Simons term, have neglected the baryonic contribution to it. The main purpose of this paper is to explore the detailed consequences of taking the U$_\textrm{Y}$(1) Chern-Simons term into account. To be more precise, we compare the simultaneous evolutions of matter asymmetries and hypermagnetic fields with and without taking the Chern-Simons term into account. Moreover, we explore the consequences of including the contributions of baryonic chemical potentials to this term, along side with the usual leptonic contributions. To accomplish this task, we choose the simple model presented in Ref.\ \cite{Dvornikov2011} and used in our previous work \cite{shiva} with the aforementioned simplifying assumptions, and use it again as a testing ground which permits us to focus on our main goal. Indeed, including other processes such as the weak sphalerons affects the results and therefore prevents us to identify and focus on the effects of our desired terms. 
%Therefore, all of our main assumptions should again be the same. 
We solve the set of coupled differential equations for the baryon and the first-generation lepton asymmetries, and the hypermagnetic field for various ranges of initial conditions in the temperature range $100$GeV$\leq T \leq10$TeV, and wherever possible compare the results with those of our previous study.

The outline of our paper is the following. In Section \ref{Static Chern-Simons Terms}, we obtain a simplified form for the coefficient of the U$_\textrm{Y}$(1) Chern-Simons term containing the baryon and the first generation lepton chemical potentials.
%, neglecting the weak sphaleron processes and their consequences.
%use the effective Euclidean Lagrangian of the gauge fields at finite fermionic density to derive the Chern-Simons coefficients in terms of the chemical potentials of right- and left-handed particle species. 
%In section \ref{Anomalous MHD Equations}, we use the flat space effective Lagrangian for the hypercharge field $Y_\mu$ containing the exact hypermagnetic Chern-Simons term of section \ref{Static Chern-Simons Terms}, to derive the dynamical equations for hypercharge fields. Then, we combine these equations to get the evolution equation of long range hypermagnetic field. In section \ref{Equilibrium In The Symmetric Phase Of The Electroweak Plasma}, we discuss the equilibrium conditions in the symmetric phase of the primordial plasma and comment on the relevant chemical potentials. 
In Sections \ref{Kinetics Of Leptons In Hypermagnetic Fields} and \ref{Baryon Asymmetry Generation In Hypermagnetic Fields}, we derive the dynamical equations for the hypermagnetic field as well as the baryon and the first generation lepton asymmetries by considering the Abelian anomaly equations and the inverse Higgs decay processes, and using the simplified coefficient of the U$_\textrm{Y}$(1) Chern-Simons term obtained in Section \ref{Static Chern-Simons Terms}. 
%In section \ref{Baryon Asymmetry Generation In Hypermagnetic Fields}, We derive the dynamical equation of the baryon asymmetry of the Universe (BAU) in hypermagnetic fields. 
In Section \ref{Discussion}, we solve the set of coupled differential equations for fermion asymmetries and the hypermagnetic field 
%in the symmetric phase down to the electroweak phase transition (EWPT) time $t_{EW}$ 
numerically and display the results. 
%We use the important conservation law $B/3-L_e=const$ merely as a consistency check on our results. 
We also use the conventions discussed in Appendix A of Ref.\ \cite{Long}, and the anomaly equations summarized in Appendix B of that reference. In Section \ref{Summary and Discussion} we summarize the results and state our conclusions.
 
\section{Static Chern-Simons Terms}\label{Static Chern-Simons Terms}
In the static limit, one can use the method of Dimensional Reduction to obtain the effective action for the soft $\textrm{SU}(2)_{\textrm{L}}$ and U$_\textrm{Y}$(1) gauge fields in which the Chern-Simons terms $c_E n_{CS}$ and $c'_E n'_{CS}$ emerge, respectively \cite{Ginsparg,Kajantie1995}. Here, the Chern-Simons densities $n_{CS}$ and $n'_{CS}$ are given by \cite{Laine}
\be\begin{split}\label{n'_CS}
n_{CS} = \frac{g^2}{32\pi^2} {\epsilon}_{ijk} (A_i^a G_{jk}^a - \frac{g}{3} f^{abc} A_i^a A_j^b A_k^c),\cr
n'_{CS} = \frac{g'^2}{32\pi^2} {\epsilon}_{ijk} Y_i Y_{jk},\ \ \ \ \ \ \ \ \ \ \ \ \ \ \ \ \ \ \ \ \ \ \ \ \ 
% = \frac{g'^2}{32\pi^2}(2\textbf{Y}.\textbf{B}_\textbf{Y}). 
\end{split}\ee 
%In the above equation, $g'$ is the U$_\textrm{Y}$(1) gauge coupling, $\textbf{Y}$ is its corresponding vector potential and $\textbf{B}_\textbf{Y}$ = $\nabla\times\textbf{Y}$ is the hypermagnetic field.
where $A_{\mu}^a$ and $Y_{\mu}$ are the $\textrm{SU}(2)_{\textrm{L}}$ and $\textrm{U}_{\textrm{Y}}(1)$ gauge fields, and $G_{\mu\nu}^a$, $Y_{\mu\nu}$, $g$ and $g'$ are their relevant field strength tensors and gauge couplings.

Let us define the notations needed in the expressions for $c_E$ and $c'_E$. Since the non-Abelian gauge interactions are in thermal equilibrium at all temperatures of interest \cite{Long}, they produce a strong driving force to equalize the asymmetries carried by different components of a given multiplet. Therefore, we can let $\mu_{Q_i}$ denote the common chemical potential of up and down left-handed quarks with different colors, $\mu_{{u_R}_i}$ ($\mu_{{d_R}_i}$) the common chemical potential of right-handed up (down) quarks with different colors, $\mu_{L_i}$($\mu_{R_i}$) the common chemical potential of left-handed (right-handed) leptons, where \lq{\textit{i}}\rq\ is the generation index. Then, the general forms of $c_E$ and $c'_E$ as given by Eqs.\ (2.4) and (2.7) of our previous study are \cite{shiva}  
\be\begin{split}\label{c'_E}
c_E = {\sum}_{i=1}^{n_G}(3\mu_{Q_i}+{\mu}_{L_i}),\ \ \ \ \ \ \ \ \ \ \ \ \ \ \ \ \ \ \ \ \ \ \ \ \ \ \ \ \ \ \ \ \ \ \ \ \ \cr
c'_E = {\sum}_{i=1}^{n_G}\left[-2\mu_{R_i} + \mu_{L_i} - \frac{2}{3}\mu_{{d_R}_i} - \frac{8}{3}\mu_{{u_R}_i} + \frac{1}{3}\mu_{Q_i}\right],
\end{split}\ee
where $n_G$ is the number of generations.

As mentioned in Section \ref{Introduction}, the simultaneous evolution of matter asymmetries and hypermagnetic fields has been studied in some of the previous works. However, the U$_\textrm{Y}$(1) Chern-Simons term has been either completely neglected or only the contribution of the first generation leptonic chemical potentials to $c'_E$ (as given by Eq.\ (\ref{c'_E})) been taken into account and that of the baryonic ones been neglected. These are precisely the issues that we want to explore in this paper, namely the consequences of considering the U$_\textrm{Y}$(1) Chern-Simons term and also its baryonic contribution.
%baryonic and the first generation leptonic asymmetries and the hypermagnetic field via the models in which the contribution of the first generation leptonic chemical potentials to $c'_E$ (as given by Eq.\ (\ref{c'_E})) is considered, but that of the baryonic ones is not taken into account \cite{Dvornikov2011,Dvornikov,Smirnov,Sokoloff,shiva}. This is precisely what we want to explore in this paper, namely the consequences of considering the baryonic contribution to $c'_E$. 
For this purpose, we choose the simple model used in our previous work \cite{shiva} as a testing ground, along with all of its simplifying assumptions including the neglect of the weak sphaleron processes. These processes, whose properties are well studied in the absence of the hypermagnetic fields, have very high rates in the symmetric phase \cite{Burnier} which keeps them in thermal equilibrium and leads to vanishing of $c_E$ as given by Eq.\ (\ref{c'_E}) (see Table 1 of Ref.\ \cite{Long}). This puts a constraint on the chemical potentials and strongly affects the scope of the aforementioned effects that we want to study. Therefore, the inclusion of weak sphalerons in the model\footnote{In order to properly include the effects of weak sphalerons in the presence of hypermagnetic fields, one can include the term corresponding to the weak sphalerons in the evolution equations of left-handed fermion asymmetries and let $c_E$ evolve freely in accordance with the evolution of its constituents as given by Eq.\ (\ref{c'_E}). When we do this for the model under study, we find that $c_E$ stays very close to zero in the whole interval; albeit near the EWPT the effect of the hypermagnetic fields via the Abelian anomalous effects becomes strong enough to force the system slightly out of equilibrium. Although the departure of $c_E$ from zero is small in this case, its consequences are non-negligible. However, the extent of this effect is model dependent. We plan to report on the results of our study which we just alluded to and includes weak sphalerons and other comparable effects, elsewhere.} adds an unnecessary complication which would obscure our results. Hence, the chosen simple model is a proper testing ground, to which we now return.

The expression for $c'_E$ as given by Eq.\ (\ref{c'_E}) can be simplified by considering the fast processes operating on the quarks. Assuming that the rates of all Yukawa interactions for quarks (up-type Yukawa in processes $d_L^i+\phi^{(+)} \leftrightarrow u_R^i$ and $u_L^i+\phi^{(0)} \leftrightarrow u_R^i$; down-type Yukawa in processes $u_L^i \leftrightarrow \phi^{(+)}+ d_R^i$ and $d_L^i \leftrightarrow \phi^{(0)} + d_R^i$, and their conjugate reactions \cite{Long}) are much higher than the Hubble expansion rate, we obtain
\be
\mu_{{u_R}_i}-\mu_{Q_i}=\mu_0,\ \ \ \ \ \mu_{{d_R}_i}-\mu_{Q_i}=-\mu_0,
\ee
where, $\mu_0$ is the chemical potential of the Higgs field. Let us assume that all up or down quarks belonging to different generations with distinct handedness have the same chemical potential (i.e., $\mu_{{u_R}_i}=\mu_{u_R},\ \mu_{{d_R}_i}=\mu_{d_R},\ \mu_{Q_i}=\mu_Q;\ i=1,2,3$) due to the flavor mixing in the quark sector (see Section 3 of the third paper of Ref.\ \cite{Campbell}). Then, we obtain
\be
\mu_{u_R}-\mu_Q=\mu_0,\ \ \ \ \ \mu_{d_R}-\mu_Q=-\mu_0,
\ee 
Since, we have the simplifying assumption of zero Higgs asymmetry, we get
\be\label{equality}
\mu_{u_R}=\mu_{d_R}=\mu_Q.
\ee
In other words, assuming zero Higgs asymmetry, the fast processes tend to equalize all quark chemical potentials. Using Eq.\ (\ref{equality}), we can simplify Eq.\ (\ref{c'_E}) in the form 
\be \label{c'_E-simp}
c'_E = {\sum}_{i=1}^{n_G}\left[-2\mu_{R_i} + \mu_{L_i} - 3\mu_Q \right].
\ee
Recalling that $N_c = 3$ and $N_w = 2$ are the ranks of non-Abelian gauge groups and $n_G=3$ is the number of generations, the whole baryonic chemical potential can be calculated as 
\be\label{muBmuQ}  
\mu_B = \frac{1}{N_c}{\sum}_{i=1}^{n_G}\left[N_cN_w\mu_{Q_i} +N_c\mu_{u_{R_i}} +N_c\mu_{d_{R_i}}\right]=12\mu_Q.
\ee
Therefore, the simplified form of $c'_E$ in terms of the baryonic and the first generation leptonic chemical potentials takes the form
%We substitute $\mu_Q$ from Eq.\ (\ref{muBmuQ}) into Eq.\ (\ref{c'_E-simp}) to obtain
\be \label{c'_E2}
c'_E = -2\mu_{e_R} + \mu_{e_L} - \frac{3}{4}\mu_B.
\ee 

\section{The Evolution Equation for the Hypermagnetic Field }\label{Kinetics Of Leptons In Hypermagnetic Fields}
Let us recall the generalized diffusion equation for the hypermagnetic field derived from the AMHD equations in our previous work (Eq.\ (3.6) of Ref.\ \cite{shiva}),
%Now, we obtain the evolution equation for the hypermagnetic field by recalling Eq.\ (3.6) of Ref.\ \cite{shiva} which is the generalized diffusion equation derived from the AMHD equations in our previous work \cite{shiva},
%Let us recall the equation of motion of the hypermagnetic field derived from the AMHD equations in our previous work \cite{shiva},
\be \label{hypermagnetic}
\frac{\partial \textbf{B}_\textbf{Y}} {\partial t} = \frac{1}{\sigma}\nabla^2\textbf{B}_\textbf{Y} + \alpha_Y\nabla\times\textbf{B}_\textbf{Y},\ \ \ \ \ \textrm{where}\ \ \alpha_Y(T) = - c'_E\frac{g'^2}{8\pi^2\sigma}. 
\ee
In the above equation, $\sigma\sim100T$ is the hyperconductivity of the plasma \cite{Arnold}, and $c'_E$ is given by Eq.\ (\ref{c'_E2}).
Choosing the simplest nontrivial configuration of the hypermagnetic field, which is
\be\label{wave}
Y_x=Y(t)\sin k_0z,\ \ \ \ \ Y_y=Y(t)\cos k_0z,\ \ \ \ \ Y_z =Y_0 =0,
\ee
%one obtains the hypermagnetic field amplitude $B_Y(t)=k_0Y(t)$. 
%We substitute $c'_E$ from Eq.\ (\ref{c'_E2}) into the expression for $\alpha_Y$ as given by Eq.\ (\ref{hypermagnetic}) and obtain
%and adhere to our simple model with the assumptions mentioned in the Introduction. That is, we consider only the contribution of baryonic and the first generation leptonic chemical potentials to $c'_E$ to be able to compare our results with those of Ref.\ \cite{shiva} and investigate the effect of baryonic contribution to $c'_E$ on the evolutions. Then, we obtain 
%\be\begin{split}\label{alpha_Y} 
%\alpha_Y(T) \simeq \left[2\mu_{e_R} - \mu_{e_L}+\frac{3}{4}\mu_B\right]\frac{g'^2}{8\pi^2\sigma}.\ \ \ \ \ \ \ \ \ \ \ \ \ \ \   
%\end{split}\ee
%\be\begin{split}\label{alpha_Y} 
%\alpha_Y(T) = -{\sum}_{i=1}^{n_G}\left[-2\mu_{R_i} + \mu_{L_i} - \frac{\mu_B}{4}\right] \frac{g'^2}{8\pi^2\sigma}\ \ \ \ \ \cr 
% \simeq \left[2\mu_{e_R} - \mu_{e_L}+\frac{3}{4}\mu_B\right]\frac{g'^2}{8\pi^2\sigma}.\ \ \ \ \ \ \ \ \ \ \ \ \ \ \   
%\end{split}\ee
and using it in Eq.\ (\ref{hypermagnetic}), one obtains the evolution equation for the hypermagnetic field amplitude $B_Y(t)=k_0Y(t)$ in the form
%Substituting $\alpha_Y(t)$ from Eq.\ (\ref{alpha_Y}) and the simple wave configuration given by Eq.\ (\ref{wave}) into Eq.\ (\ref{hypermagnetic}), we obtain the evolution equation for $B_Y(t)$ in the form
\be \label{B_Y(t)}
\frac{dB_Y}{dt} = B_Y\left[-\frac{k_0^2}{\sigma}+\frac{k_0 g'^2}{4\pi^2\sigma}\left(\mu_{e_R} - \frac{\mu_{e_L}}{2}+\frac{3}{8}\mu_{B}\right)\right].
\ee
In the above equation, the coupling of the evolution of the hypermagnetic field to those of the chemical potentials is apparent. In the next section we discuss the latter, however let us first obtain the relevant expression for the Abelian anomaly ($\sim \textbf{E}_\textbf{Y}.\textbf{B}_\textbf{Y}$) appearing in the dynamical equations of the fermionic asymmetries.
%fermionic sector.

Let us recall the generalized Ohm's law derived from the AMHD equations in our previous work (Eq.\ (3.4) of Ref.\ \cite{shiva}),
%Let us recall Eq.\ (3.4) of Ref.\ \cite{shiva}, which is the generalized Ohm's law derived from the AMHD equations,
\be\label{E_Y}
\textbf{E}_\textbf{Y} = - \textbf{V}\times\textbf{B}_\textbf{Y} + \frac{\nabla\times\textbf{B}_\textbf{Y}}{\sigma} - \alpha_Y\textbf{B}_\textbf{Y}. 
\ee
Using the above equation with $\sigma = 100T$, and $\alpha_Y$ and $c'_E$ as given by Eqs.\ (\ref{hypermagnetic}) and (\ref{c'_E2}), for the simple configuration of the hypermagnetic field given by Eq.\ (\ref{wave}), the form of the Abelian anomaly simplifies to
%\be\label{E_Y.B_Y2} 
%\textbf{E}_\textbf{Y}.\textbf{B}_\textbf{Y} = \frac{1}{\sigma}(k_0B_Y^2) - \alpha_Y B_Y^2.
%\ee
%Substituting the expression for $c'_E$ given by Eq.\ (\ref{c'_E2}) into Eq.\ (\ref{E_Y.B_Y2}) and using $\sigma = 100T$, we obtain 
\be\label{E_Y.B_Y3}
\textbf{E}_\textbf{Y}.\textbf{B}_\textbf{Y} = \frac{B_Y^2}{100} \left[\frac{k_0}{T}-\frac{g'^2}{4\pi^2T}\left(\mu_{e_R}-\frac{\mu_{e_L}}{2}+\frac{3}{8}\mu_B\right)\right].
\ee

\section{The Dynamical Equations for the Lepton and Baryon Asymmetries}\label{Baryon Asymmetry Generation In Hypermagnetic Fields}
%Let us consider the electroweak plasma slightly out of thermal equilibrium due to the presence of the large scale hypermagnetic fields or some unbalanced chemical potentials. We mentioned in Section \ref{Introduction} that the hypercharge fields are coupled to the fermionic number densities because of the anomaly. We set up and solve the coupled system of Boltzmann-type equations for these chemical potentials and the hypermagnetic fields. 

%We obtain the evolution equations for leptonic asymmetries by considering the Abelian anomalous contributions and electron chirality flip through inverse Higgs decay processes. The violation of lepton numbers via the Abelian anomaly is given by \cite{Long},

In the Standard Model, the U$_\textrm{Y}$(1) Abelian anomaly violates the first generation lepton numbers in the following form: 
\bea\label{e_RL}
\partial_\mu j_{e_R}^\mu &= -\frac{1}{4}(Y_R^2) \frac{g'^2}{16\pi^2}Y_{\mu\nu} {\tilde{Y}}^{\mu\nu} &= \frac{g'^2}{4\pi^2}(\textbf{E}_\textbf{Y}.\textbf{B}_\textbf{Y}),\\*\nonumber
\partial_\mu j_{\nu_e^L}^\mu = \partial_\mu j_{e_L}^\mu &= +\frac{1}{4}(Y_L^2) \frac{g'^2}{16\pi^2}Y_{\mu\nu} {\tilde{Y}}^{\mu\nu} &= - \frac{g'^2}{16\pi^2}(\textbf{E}_\textbf{Y}.\textbf{B}_\textbf{Y}),
\eea
where ${\tilde{Y}}^{\mu\nu}$ is the dual field strength tensor, and the relevant hypercharges are $Y_R = -2$ and $Y_{L} = -1$. Therefore, the system of dynamical equations for the corresponding asymmetries, taking into account the Abelian anomaly Eqs.\ (\ref{e_RL}) and inverse Higgs decay processes, takes the form\footnote{We have used Appendix B of Ref.\ \cite{Dvornikov} but with the assumption of zero Higgs asymmetry. See also Eq.\ (2.6) in Section 2.1 of Ref.\ \cite{Kamada} for the general form of the equations.}
\bea\label{lepton equations} 
\frac{d\eta_{e_R}}{dt} &= +\frac{g'^2}{4\pi^2s}(\textbf{E}_\textbf{Y}.\textbf{B}_\textbf{Y}) + 2\Gamma_{RL}(\eta_{e_L}-\eta_{e_R}),\\*\nonumber
\frac{d\eta_{\nu_e^L}}{dt} =\frac{d\eta_{e_L}}{dt} &= -\frac{g'^2}{16\pi^2s}(\textbf{E}_\textbf{Y}.\textbf{B}_\textbf{Y}) + \Gamma_{RL}(\eta_{e_R}-\eta_{e_L}).
%\frac{d\eta_{\nu_e^L}}{dt}& = -\frac{g'^2}{16\pi^2s}(\textbf{E}_\textbf{Y}.\textbf{B}_\textbf{Y}) + \Gamma_{RL}(\eta_{e_R}-\eta_{\nu_e^L}).
\eea
In Eqs.\ (\ref{lepton equations}), $\eta_f = (n_f -n_{\bar{f}})/s$ with $f=\{e_R,e_L,\nu_e^L\}$ is the matter asymmetry, $s=2\pi^2g^*T^3/45$ is the entropy density of the Universe and $g^*=106.75$ is the number of relativistic degrees of freedom. $\Gamma_{RL}$ is the rate of inverse Higgs decay reactions, and the factor 2 multiplying it in the first line is because of the equivalent rates of reaction branches ($e_L \bar{e}_R\rightarrow\phi^{(0)}$ and $\nu_e^L \bar{e}_R\rightarrow\phi^{(+)}$ and their conjugate processes). Since the SU(2) gauge interactions are very fast, $\eta_{e_L}\approx \eta_{\nu_e^L}$ and the evolution equation of the neutrino asymmetry is unnecessary.

Let us define the variable $x = t/t_{EW} = (T_{EW}/T )^2$, in accordance with the Friedmann law, where $t_{EW} = M_{Pl}^*/2T_{EW}^2$ and $M_{Pl}^* = M_{\textrm{Pl}}/1.66\sqrt{g^*}$ is the reduced Planck mass. Then, $\Gamma_{RL} = \Gamma_0(1-x)/2t_{EW}\sqrt{x}$ with $\Gamma_0 = 121$ \cite{Campbell,Dvornikov2011,Dvornikov}. 
%the variable $x = t/t_{EW} = (T_{EW}/T )^2$ in accord with the Friedmann law, $t_{EW} = M_0/2T_{EW}^2$ and $M_0 = M_{\textrm{Pl}}/1.66\sqrt{g^*}$.
Recalling the equation $n_f -n_{\bar{f}} = \mu_f T^2/6$ for fermions, and defining $y_f = 10^4\mu_f/T$, the fermion asymmetry will be $\eta_f = 10^{-4}y_f T^3/6s$. Using Eq.\ (\ref{E_Y.B_Y3}), Eqs.\ (\ref{lepton equations}) can be rewritten in terms of the variables $y_f$ in the following form 
%Using (\ref{E_Y.B_Y3}) and defining $y_b(x) = 10^4\xi_{b}(x)$, the equations (\ref{asymmetry equations}) can be rewritten in the form
\be\begin{split}\label{y_equations}
\frac{dy_R}{dx} = \left[B_0 x^{1/2}-A_0y_T\right]\left(\frac{B_Y(x)}{10^{20}\mbox{G}}\right)^2x^{3/2} -\Gamma_0\frac{1-x}{\sqrt{x}}(y_R-y_L),\ \ \ \ \ \cr
\frac{dy_L}{dx} = \frac{-1}{4}\left[B_0 x^{1/2}-A_0y_T\right]\left(\frac{B_Y(x)}{10^{20}\mbox{G}}\right)^2x^{3/2} -\Gamma_0\frac{1-x}{2\sqrt{x}}(y_L-y_R),
%B_0 = 25.6\left(\frac{k_0}{10^{-7}T_{EW}}\right),\ \ \ A_0 = 77.6,\ \ \ y_T=y_R-\frac{y_L}{2}+\frac{137}{108}y_B.
\end{split}\ee
where,
\be\label{A0B0}
B_0 = 25.6\left(\frac{k_0}{10^{-7}T_{EW}}\right),\ A_0 = 77.6,\ \textrm{and}\ \ y_T=y_R-\frac{y_L}{2}+\frac{3}{8}y_B.
\ee
We have chosen the overall scale of $B_0$ and $A_0$ to normalize the hypermagnetic field at $10^{20}$G.

In the Standard Model, the anomalous processes change the baryon asymmetry $\eta_B = (n_B-n_{\bar{B}})/s$ as well as the lepton asymmetry of each generation $\eta_{L_i} = (n_{L_i}-n_{\bar{L}_i})/s$, respecting the conservation law $\eta_B/3 - \eta_{L_i} = \mbox{constant}$. Using this fact for the first generation asymmetries, one can obtain the evolution equation of the baryon asymmetry in the form,
\be\label{yB}
\frac{1}{3}\frac{d\eta_B}{dt} = \frac{d\eta_{e_R}}{dt} + \frac{d\eta_{e_L}}{dt} + \frac{d\eta_{\nu_e^L}}{dt},\ \ \ \textrm{or}\ \ \ \frac{1}{3}\frac{dy_B}{dx} = \frac{dy_R}{dx}+ 2\frac{dy_L}{dx},
\ee
%or equivalently in terms of the variables $y_b$,
%\be\label{yB} 
%\frac{1}{3}\frac{dy_B}{dx} = \frac{dy_R}{dx}+ 2\frac{dy_L}{dx}. 
%\ee
where $y_B=4\times10^4\pi^2g^*\eta_B/15$ is the scaled baryon asymmetry. Finally, we use Eqs.\ (\ref{yB}) and (\ref{y_equations}) to obtain,
\be\label{yB_equation}
\frac{dy_B}{dx} = \frac{3}{2}\left[B_0 x^{1/2}-A_0\left(y_R-\frac{y_L}{2}+\frac{3}{8}y_B\right)\right]\left(\frac{B_Y(x)}{10^{20}G}\right)^2x^{3/2}.
\ee
%It should be noted that $B=(n_B-n_{\bar{B}})/s=\mu_BT^2/6s=\xi_BT^3/6s=10^{-4} y_B T^3/6s=45 \times 10^{-4} y_B/12\pi^2 g^*$.
We also rewrite Eq.\ (\ref{B_Y(t)}) in terms of $x$ and the variables $y_f$ to obtain
\be \label{B_Y(x)}
\frac{dB_Y}{dx} = 3.5\left(\frac{k_0}{10^{-7}T_{EW}}\right)\left[\frac{y_T}{\pi}-0.1\left(\frac{k_0}{10^{-7}T_{EW}}\right)\sqrt{x})\right]B_Y(x),
\ee
where $y_T$ is given by Eq.\ (\ref{A0B0}). 
%The initial conditions for our first investigation chosen at $x_0 = 10^{-4}$, which is equivalent to $T_0 = 10$TeV, are
%\be
% y_R^{(0)}=y_R(x_0) = 10^{-6},\ \ \ \ \ y_L^{(0)}=y_L(x_0) = 0.
%\ee
%Such conditions correspond to the right-handed electron asymmetry $\xi_{e_R}(x_0) = 10^{-10}$  \cite{Dvornikov2011,Dvornikov}.

%\section{Results}\label{Discussion}
%\begin{figure} 
%  \includegraphics[width=65mm]{etaBetaRetaLplot.pdf}
%  \hspace{2mm}
%  \includegraphics[width=65mm]{Byplot.pdf}
%\caption{The time plots of (left): baryon asymmetry $\eta_B$ (dashed line), right-handed electron asymmetry $\eta_R=\eta_{e_R}$ (dotted line), left-handed lepton asymmetry $\eta_L=\eta_{e_L}=\eta_{\nu_e^L}$ (solid line) and (right): hypermagnetic field amplitude $B_Y$, for $k_0=10^{-7} T_{EW}$, $y_R^{(0)} = 10^{-6}$ and $B_Y^{(0)}=10^{21}G$. The maximum relative error for these plots is of the order of $10^{-18}$.
%}\label{RGB}
%\end{figure}

\section{Results}\label{Discussion} 
\begin{figure} 
%\begin{center}
  \includegraphics[width=65mm]{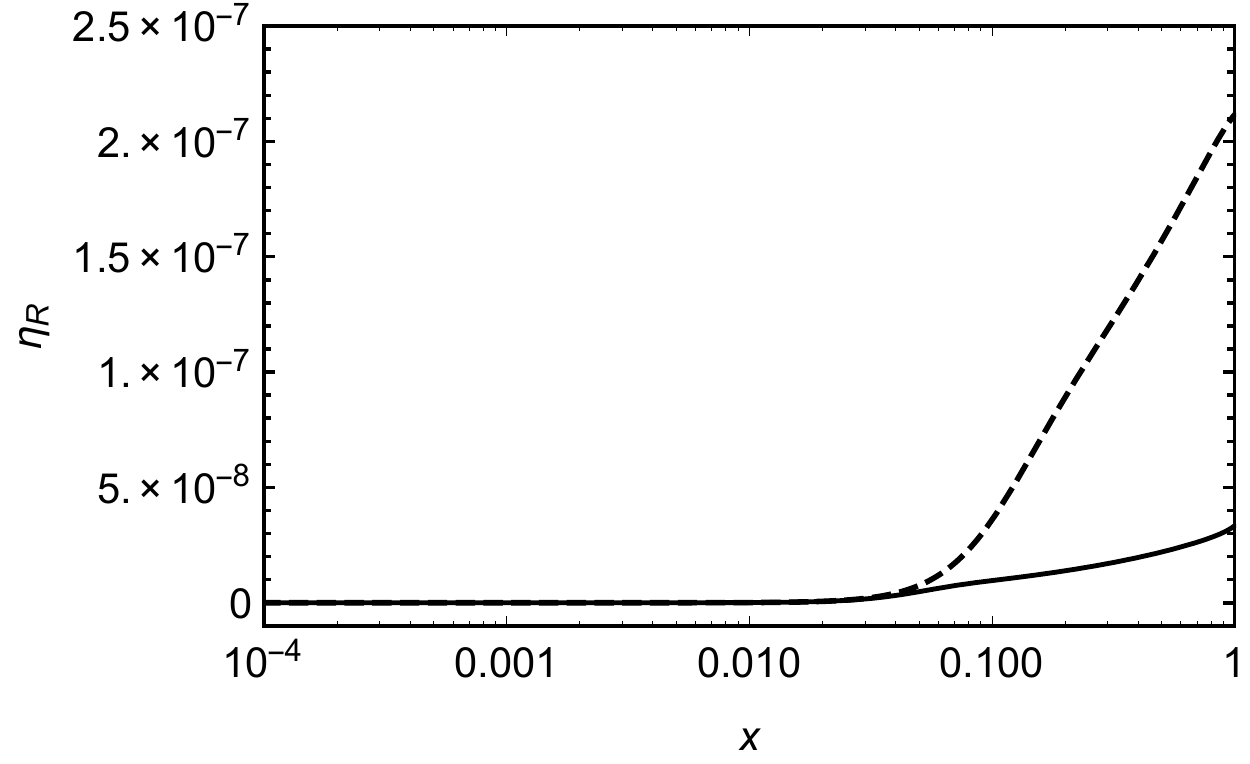}
  \hspace{2mm}
  \includegraphics[width=65mm]{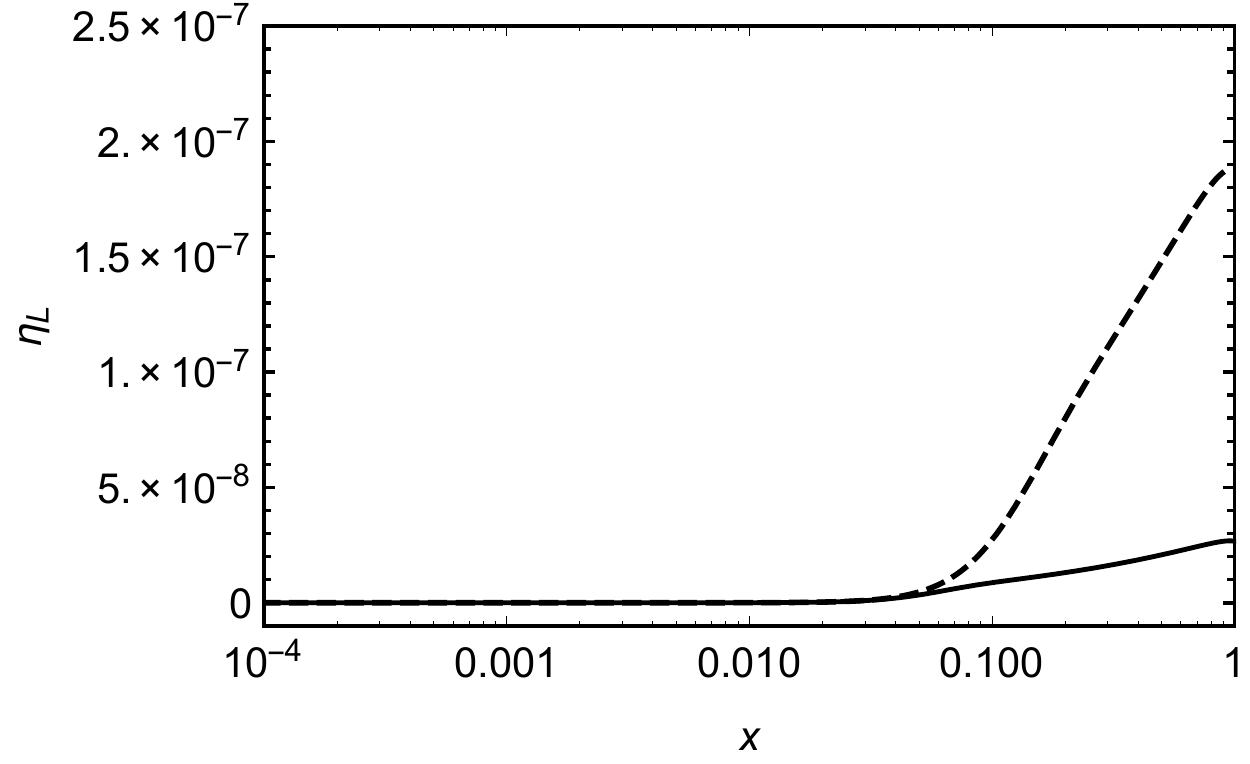}
%\caption{ $E_0=0.001$ and $k=0.5$(Left) and $k=1.5$(Right) }\label{1Jt}
%\end{center}
%\end{figure}
%\begin{figure}
%\begin{center}
  \includegraphics[width=65mm]{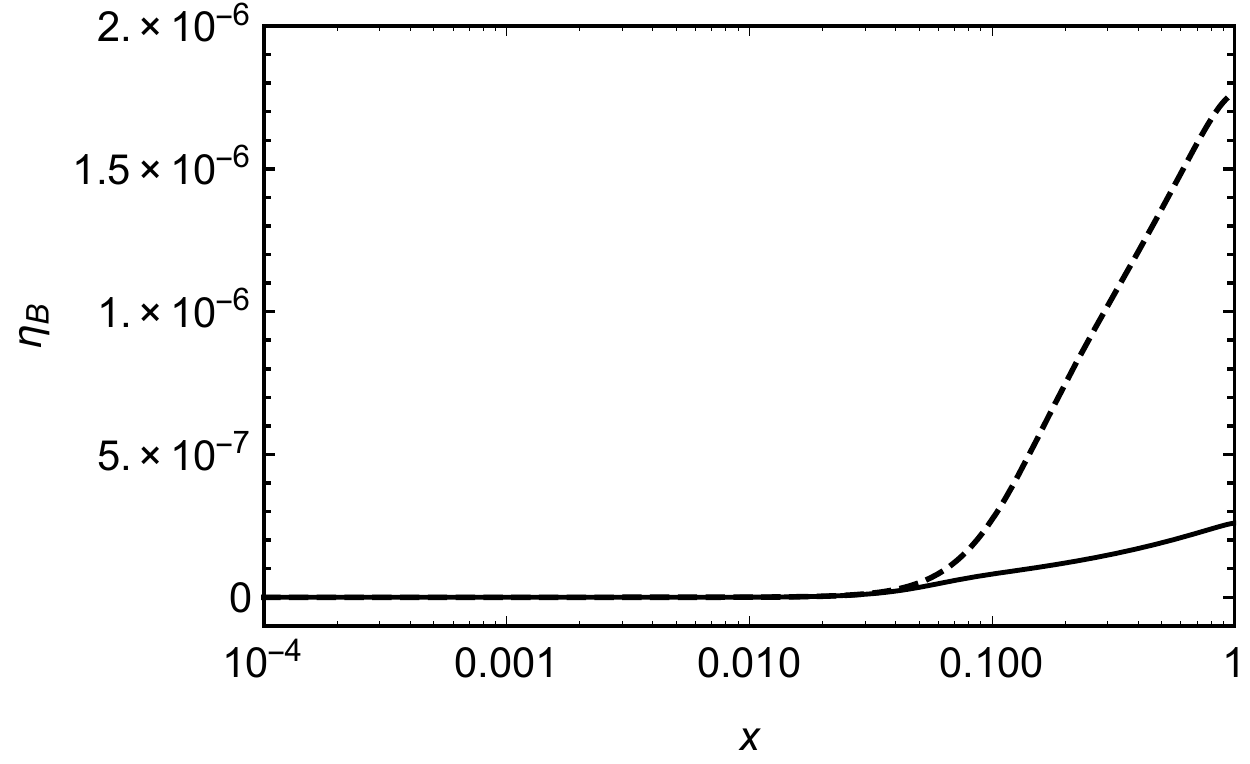}
  \hspace{2mm}
  \includegraphics[width=65mm]{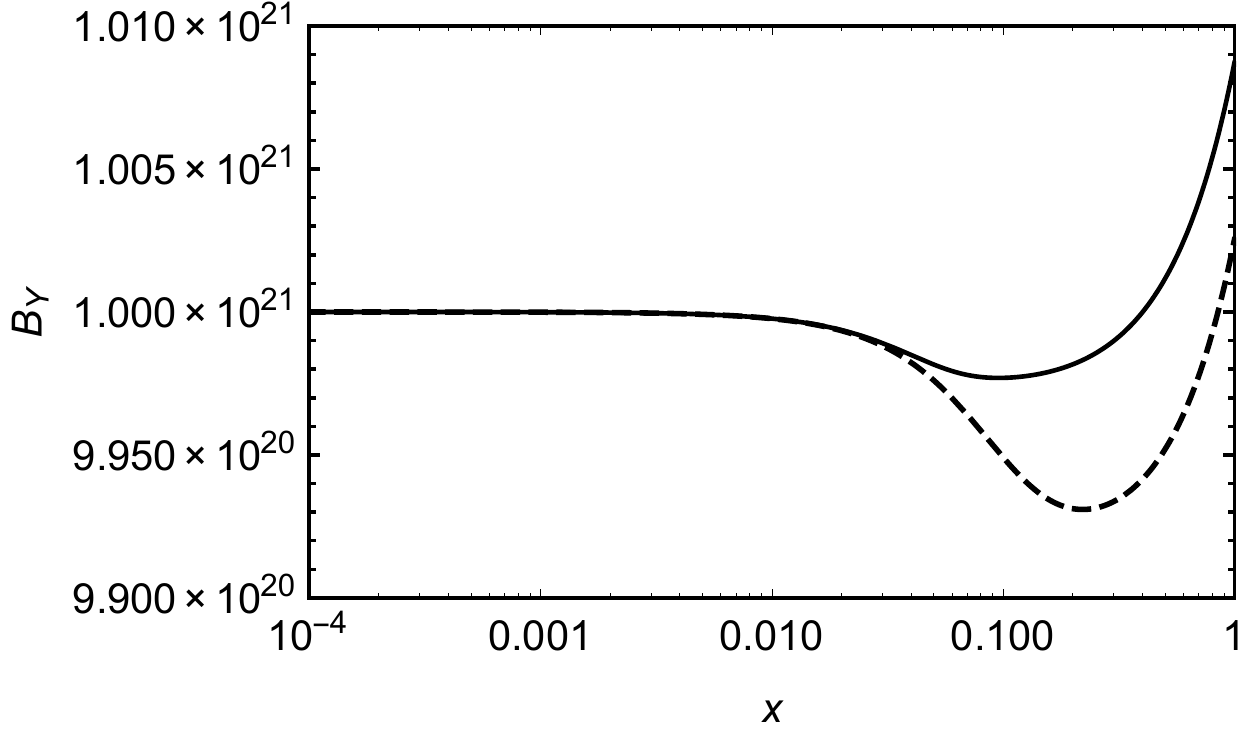}
\caption{The time plots of the first-generation leptonic asymmetries $\eta_R=\eta_{e_R}$ and $\eta_L=\eta_{e_L}=\eta_{\nu_e^L}$, baryonic asymmetry $\eta_B$, and the hypermagnetic field amplitude $B_Y$ for $k_0=10^{-7} T_{EW}$ with initial conditions $B_Y^{(0)}=10^{21}G$ and zero initial matter asymmetries for two different cases.\\
 Case 1 (dashed lines): \ \ \  $\alpha_Y^{(0)}=\frac{g'^2}{8\pi^2\sigma}(2\mu_{e_R} - \mu_{e_L})$.\\ Case 2 (solid lines): \ \ \ \ \ \  $\alpha_Y=\frac{g'^2}{8\pi^2\sigma}(2\mu_{e_R} - \mu_{e_L} + \frac{3}{4}\mu_B)$.\\ 
The starting point is at $T_0=10\ $TeV, $x_0=\frac{t_0}{t_{EW}}=(\frac{T_{EW}}{T_0})^2=10^{-4}$ and the final point is at $T_f=T_{EW}$, $x_f=\frac{t_f}{t_{EW}}=(\frac{T_{EW}}{T_f})^2=1$. Case 1 is obtained from our previous work \cite{shiva} and is reproduced here for comparison. The maximum relative error for these plots is of the order of $10^{-20}$.
%top(left) :\ \ \ \ \ \ \ \ Normalized chemical potential of right-handed electrons, \\
%top(right) :\ \ \ \ \ \ Normalized chemical potential of left-handed electrons, \\ 
%bottom(left) : \ \ Baryon asymmetry, \\
%bottom(right) : Hypermagnetic field amplitude. 
}\label{RGB}
\end{figure}

The simplified form of the U$_\textrm{Y}$(1) Chern-Simons coefficient $c'_E$ is given in Eq.\ (\ref{c'_E2}) and affects the evolution equations (\ref{y_equations}), (\ref{yB_equation}), and (\ref{B_Y(x)}) through $\alpha_Y$ as given by Eq.\ (\ref{hypermagnetic}). In this section, we study the effect of the Chern-Simons term 
%and also the contribution of the baryonic chemical potentials to it 
on the evolution of matter asymmetries and hypermagnetic fields for a variety of initial conditions. 
%categorized under the headings "Matter Asymmetry Generation by Hypermagnetic Fields" and "Hypermagnetic Fields Growth by Matter Asymmetries". 
To accomplish this task, we compare the results for three different choices of $\alpha_Y$, namely $\alpha_Y^{(0)}=\frac{g'^2}{8\pi^2\sigma}(2\mu_{e_R} - \mu_{e_L})$ (absence of the baryonic contribution), $\alpha_Y=\frac{g'^2}{8\pi^2\sigma}(2\mu_{e_R} - \mu_{e_L} + \frac{3}{4}\mu_B)$ (as given by Eqs.\ (\ref{hypermagnetic}) and (\ref{c'_E2})), and $c\alpha_Y$ where $c=\{0,0.1,0.2\}$, i.e. attenuated Chern-Simons term, with a given set of initial conditions. 
%The initial conditions are categorized under the headings "Matter Asymmetry Generation by Hypermagnetic Fields" and "Hypermagnetic Fields Growth by Matter Asymmetries". 
Moreover, $k_0$ is set to $k_{max}=10^{-7} T_{EW}$ which is the maximum wave number surviving Ohmic dissipation.

%The contribution of baryonic chemical potentials to the U$_\textrm{Y}$(1) Chern-Simons term shows up in Eq.\ (\ref{c'_E2}) and affects the evolution equations (\ref{y_equations},\ref{yB_equation},\ref{B_Y(x)}) through $\alpha_Y$ as given by Eq.\ (\ref{hypermagnetic}). We study the effect of including this contribution on the evolution of matter asymmetries and hypermagnetic fields for a variety of initial conditions. Moreover, $k_0$ is set to $k_{max}=10^{-7} T_{EW}$ which is the maximum wave number surviving Ohmic dissipation.

\begin{figure} 
  \includegraphics[width=65mm]{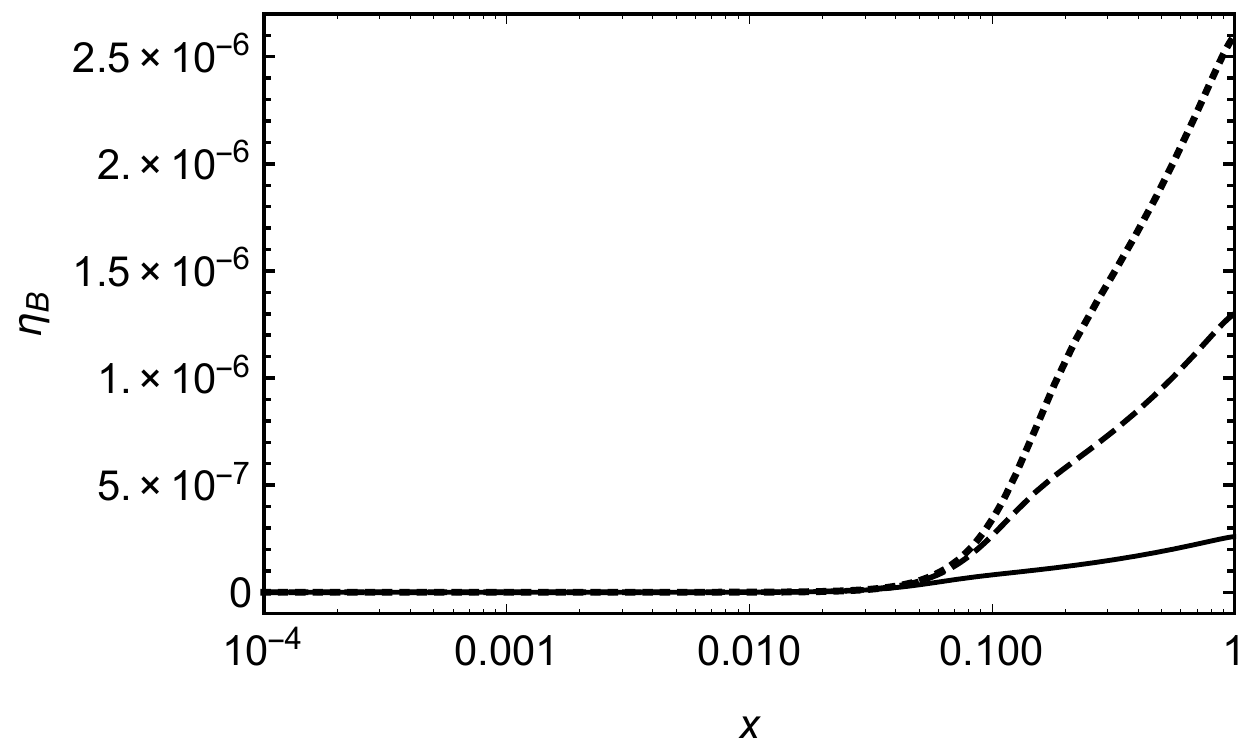}
  \hspace{2mm}
  \includegraphics[width=65mm]{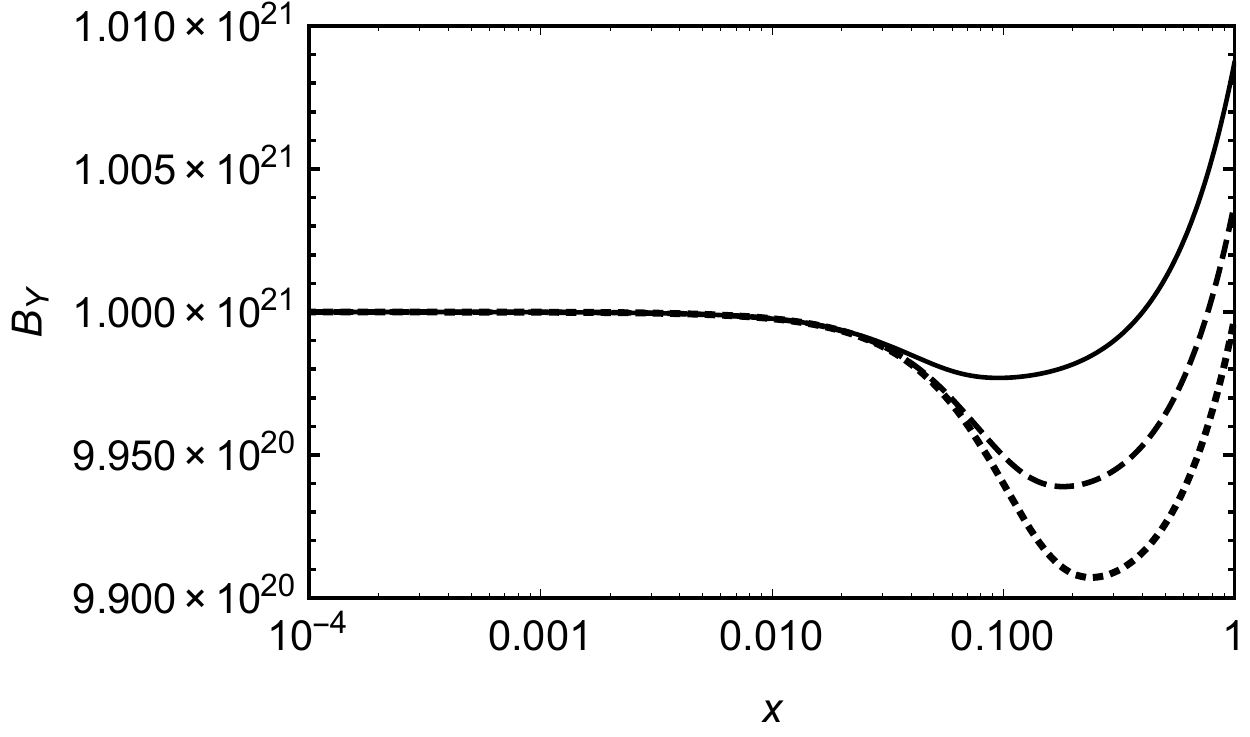}

  \includegraphics[width=65mm]{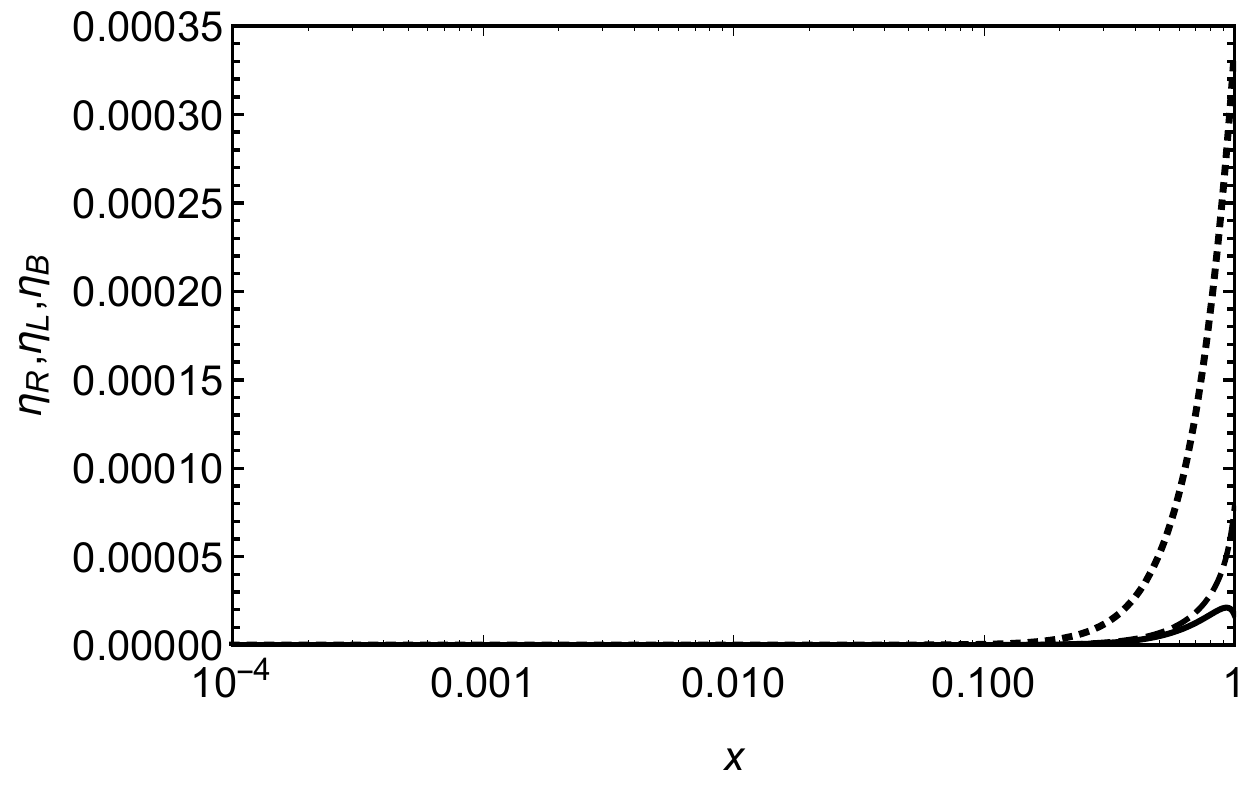}
  \hspace{2mm}
  \includegraphics[width=65mm]{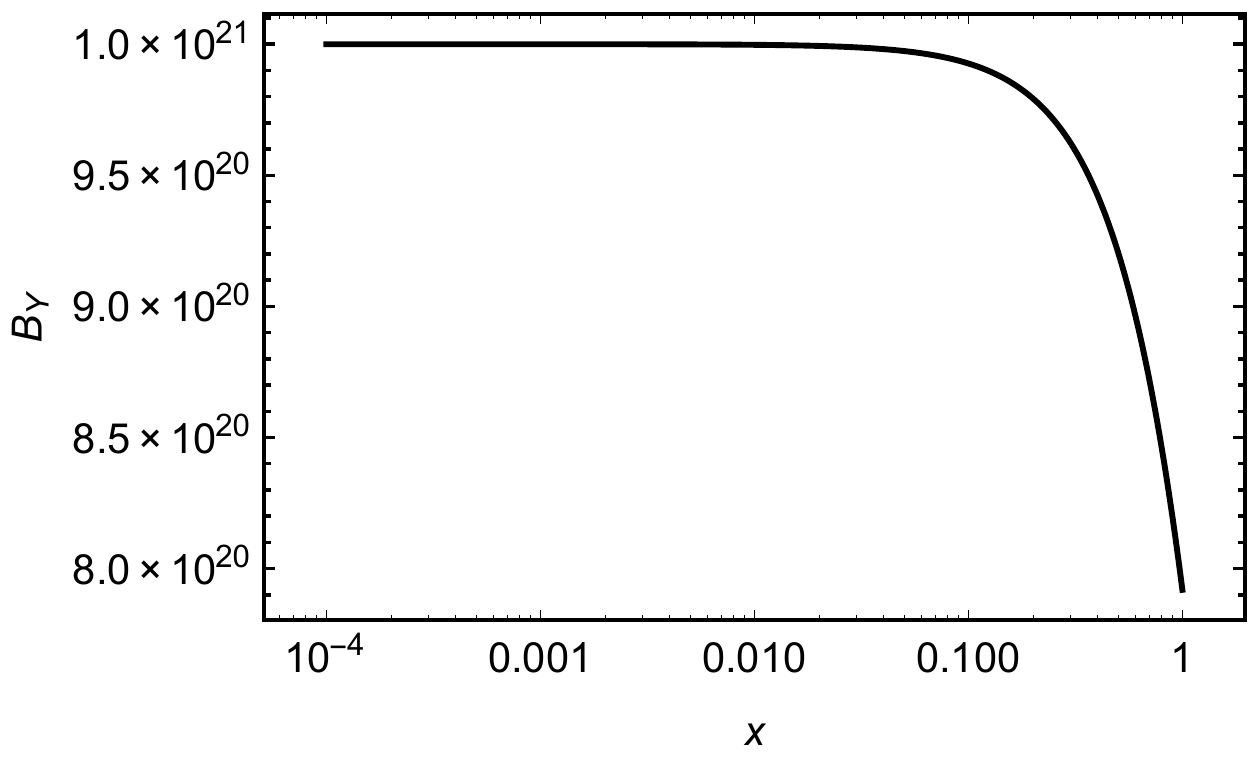}
\caption{Top figures: The time plots of baryonic asymmetry $\eta_B$ and the hypermagnetic field amplitude $B_Y$ for $k_0=10^{-7} T_{EW}$ with initial conditions $B_Y^{(0)}=10^{21}G$, zero initial matter asymmetries and attenuated hypermagnetic helicity coefficient $c\alpha_Y$ for three different values of c. That is, $c=1$ (solid lines), $c=0.2$ (dashed lines), and $c=0.1$ (dotted lines). Bottom figures: The time plots of the first-generation leptonic asymmetries $\eta_R=\eta_{e_R}$ (dashed line) and $\eta_L=\eta_{e_L}=\eta_{\nu_e^L}$ (solid line), baryonic asymmetry $\eta_B$ (dotted line), and the hypermagnetic field amplitude $B_Y$ in the absence of the U$_\textrm{Y}$(1) Chern-Simons term ($c=0$). The maximum relative error for these plots is of the order of $10^{-16}$.\\
}\label{c}
\end{figure}
%\begin{figure} 
%  \includegraphics[width=65mm]{etaRplotc.pdf}
%  \hspace{2mm}
%  \includegraphics[width=65mm]{etaLplotc.pdf}
%  \includegraphics[width=65mm]{etaBplotc.pdf}
%  \hspace{2mm}
%  \includegraphics[width=65mm]{Byplotc.pdf}
%\caption{The time plots of the first-generation leptonic asymmetries $\eta_R=\eta_{e_R}$ and $\eta_L=\eta_{e_L}=\eta_{\nu_e^L}$, baryonic asymmetry $\eta_B$ and the hypermagnetic field amplitude $B_Y$ for $k_0=10^{-7} T_{EW}$ with initial conditions $B_Y^{(0)}=10^{21}G$ and zero initial matter asymmetries for three different cases.\\
% Case 1 (solid lines): \ \ \ \ \ \ \ \  $c=1$\\ Case 2 (dashed lines): \ \ \ \ \ $c=0.2$\\Case 3 (dotted lines): \ \ \ \ \ $c=0.1$\\ 
%The maximum relative error for these plots is of the order of $10^{-18}$.
%}\label{c}
%\end{figure}

%\begin{itemize}
%\item{\textbf{Matter Asymmetry Generation by Hypermagnetic Fields}}\\
\subsection{Matter Asymmetry Generation by Hypermagnetic Fields}\label{Matter Asymmetry Generation by Hypermagnetic Fields}
%First, we assume an strong hypermagnetic field with the initial amplitude of $B_Y^{(0)}=10^{21}$G, but zero initial values for matter asymmetries. The evolution equations are solved numerically and the relevant time plots are presented in Figure \ref{RGB}.
%At first, we assume zero initial fermion asymmetries but the initial amplitude of the hypermagnetic field to be in the range $10^{17}$G$\ <B_Y^{(0)}<10^{22}$G. The evolution equations are solved numerically and the time plots are presented in Figure \ref{RGB} for $B_Y^{(0)}=10^{21}$G 
First, the evolution equations are solved numerically by assuming zero initial matter asymmetries but an initial amplitude of the hypermagnetic field $B_Y^{(0)}=10^{21}$G for two different cases, namely $\alpha_Y^{(0)}=\frac{g'^2}{8\pi^2\sigma}(2\mu_{e_R} - \mu_{e_L})$ and $\alpha_Y=\frac{g'^2}{8\pi^2\sigma}(2\mu_{e_R} - \mu_{e_L} + \frac{3}{4}\mu_B)$. The results are presented as time plots in Figure \ref{RGB}.
%We first assume the initial values $y_R^{(0)} = 10^{-6}$, $B_Y^{(0)}=10^{21}$G. We solve the dynamical equations numerically and present the results as time plots in Figure \ref{RGB}. 
As can be seen, in both cases, matter asymmetry generation occurs in the presence of hypermagnetic fields; 
%the hypermagnetic field is able to produce matter asymmetries, 
however, the final values of the asymmetries at the onset of EWPT for the second case are almost $7$ times smaller than those of the first case. Moreover, the hypermagnetic field amplitude behaves nearly the same with a little more increase in its final value for the second case. For the rest of this subsection we use $\alpha_Y$ as given by Eqs.\ (\ref{hypermagnetic}) and (\ref{c'_E2}), that is including the baryonic contribution.

%\begin{figure} 
%  \includegraphics[width=65mm]{etaRplotc=0.pdf}
%  \hspace{2mm}
%  \includegraphics[width=65mm]{etaLplotc=0.pdf}
%  \includegraphics[width=65mm]{etaBplotc=0.pdf}
%  \hspace{2mm}
%  \includegraphics[width=65mm]{Byplotc=0.pdf}
%\caption{The time plots of the first-generation leptonic asymmetries $\eta_R=\eta_{e_R}$ and $\eta_L=\eta_{e_L}=\eta_{\nu_e^L}$, baryonic asymmetry $\eta_B$ and the hypermagnetic field amplitude $B_Y$ for $k_0=10^{-7} T_{EW}$ with initial conditions $B_Y^{(0)}=10^{21}G$ and zero initial matter asymmetries in the absence of the U$_\textrm{Y}$(1) Chern-Simons term ($c=0$).\\
% The maximum relative error for these plots is of the order of $10^{-18}$.
%}\label{c=0}
%\end{figure}

Let us examine the importance of the U$_\textrm{Y}$(1) Chern-Simons term via attenuating its effect by multiplying it with an adjustable parameter $c\leq1$.
%for the evolution of matter asymmetries and the hypermagnetic field.
%$c'_E n'_{CS}$, where $c'_E$ and $n'_{CS}$ are given by Eqs.\ (\ref{c'_E2}) and (\ref{c'_E2}) respectively. 
%To accomplish this task, we intentionally attenuate the effect of this term by multiplying it with a variable $c<1$. 
We numerically solve the evolution equations with the aforementioned initial conditions for three different values of $c:\{0.2,0.1,0\}$ and present the results as time plots, along with the case $c=1$ obtained earlier, in Fig. \ref{c}.
%In continuation, we intentionally attenuate the effect of Abelian Chern-Simons term multiplying it with a variable $c<1$. 
Figure \ref{c} shows that the smaller the value of $c$, the larger the matter asymmetries and the weaker the hypermagnetic field at $T_{EW}$. The case $c=0$ also shows that the hypermagnetic field is able to generate substantial matter asymmetries through the Abelian anomaly even in the absence of the U$_\textrm{Y}$(1) Chern-Simons term. Therefore, taking into account the U$_\textrm{Y}$(1) Chern-Simons term leads to a severe decrease in the generated matter asymmetries but a very small increase in the strength of the hypermagnetic field, all at $T=T_{EW}$. 

Let us return to our first investigation but change $B_Y^{(0)}$ in the range $10^{17}$G$\ <B_Y^{(0)}<10^{22}$G. We have solved the equations and obtained the final values of the matter asymmetries and the hypermagnetic field amplitude at $T=T_{EW}$. We do not display the results for space limitation, and suffice it to point out the salient features of this investigation. This investigation is analogous to the one done in our previous work (Fig. 2 of Ref.\ \cite{shiva}) and the results are qualitatively similar. That is, the final asymmetries increase approximately quadratically for $B_Y^{(0)}\lesssim10^{19.5}$G and saturate for $B_Y^{(0)}\gtrsim10^{20.5}$G. However, the saturated values are about 7 times smaller than those of our previous work where we used $\alpha_Y^{(0)}$. The amplitude $B_Y$ stays relatively unchanged except for $B_Y^{(0)}\gtrsim10^{20}$G, where it increases slightly above its initial value, indicating a mild resonance effect.

Next, we repeat the above investigation in the absence of the U$_\textrm{Y}$(1) Chern-Simons term by setting c=0. Interestingly, we observe that 
%the behavior is drastically different. There is no inflection point and no resonance effect. Moreover, 
the final asymmetries again increase quadratically with increasing $B_Y^{(0)}$ due to the Abelian anomaly without any saturation. Moreover, the final value of $B_Y$ decreases slightly as compared to its initial value $B_Y^{(0)}$.
%\begin{table}[ht]
%\caption{Parameters}
%\centering
%\begin{tabular}{c c c c c c}
%\hline\hline 
%Case & $p$ & $q$ & $\alpha$ & $\beta$ & $\gamma$ \\[0.5ex]
%\hline
%1 & 3 & 7 & 4 & -5 & $1$ \\
%2 & 3 & 5 & 4 & -4 & $1$ \\
%3 & 4 & 6 & 6 & -7 & $1/64$ \\
%4 & 4 & 8 & 6 & -9 & $1/64$ \\[1ex]
%\hline
%\end{tabular}
%\end{table}

%\begin{figure} 
%  \includegraphics[width=65mm]{three-timeplot-ethaR_ethaL.pdf}
%  \hspace{2mm}
%  \includegraphics[width=65mm]{three-timeplot-ethaRp_ethaLp.pdf}
% \includegraphics[width=65mm]{three-timeplot-ethaB.pdf}
% \hspace{2mm}
% \includegraphics[width=65mm]{three-timeplot-By.pdf}
%\caption{The time plots of (top left): right-handed electron asymmetry $\eta_R=\eta_{e_R}$ (dotted line), left-handed lepton asymmetry $\eta_L=\eta_{e_L}=\eta_{\nu_e^L}$ (solid line), (top right): the zoomed view, 
%(bottom left): baryon asymmetry $\eta_B$ and (bottom right): hypermagnetic field amplitude $B_Y$, for $k_0=10^{-7}T_{EW}$, $B_Y^{(0)}=10^{-2}$G and $y_R^{(0)} = 10^3$. 
%The maximum relative error for these graphs is of the order of $10^{-14}$.
%} \label{three-timeplot}
%\end{figure}

%\item{\textbf{Hypermagnetic Fields Growth by Matter Asymmetries}}\\
\subsection{Hypermagnetic Fields Growth by Matter Asymmetries}\label{Hypermagnetic Fields Growth by Matter Asymmetries}
In continuation, we examine the possibility of producing a hypermagnetic field from initial matter asymmetries, when no initial seed of the hypermagnetic field is present in the plasma. We observe that no hypermagnetic field with simple wave configuration as given by Eq.\ (\ref{wave}) can be generated.
%In our third analysis, we examine the possibility of producing a hypermagnetic field assuming that all fermion species, except the right-handed electron, have zero initial asymmetries and no initial hypermagnetic field is present as well. We solve the dynamical equations for $y_R^{(0)} = 10^3$ and observe that, similar to our previous work, no hypermagnetic field with simple wave configuration (\ref{wave}) can be generated. Thus, no net baryon asymmetry is produced and the net lepton asymmetry remains unchanged as well. Indeed, the chirality flip processes wash out the chiral asymmetry by transforming some amounts of asymmetry from right handed electrons into left-handed leptons. 
The following integral form for the evolution equation of the hypermagnetic field amplitude (\ref{B_Y(x)}) clarifies that the amplitude stands at zero if its initial value is zero:       
\be\begin{split}\label{exact-solution}
B_Y(x)=B_Y^{(0)} \exp \left[\frac{3.5k_0}{10^{-7}T_{EW}} \int_{x_0}^x \left(\frac{y_T(x')}{\pi}-\frac{0.1k_0}{10^{-7}T_{EW}}\sqrt{x'}\right)dx'\right],\cr
\textrm{where}\ \ \ y_T(x')=y_R(x')-\frac{y_L(x')}{2}+\frac{3}{8}y_B(x').\ \ \ \ \ \ \ \ \ \ \ \ \ \ \ \   
\end{split}\ee

\begin{figure} 
%\begin{center}
  \includegraphics[width=65mm]{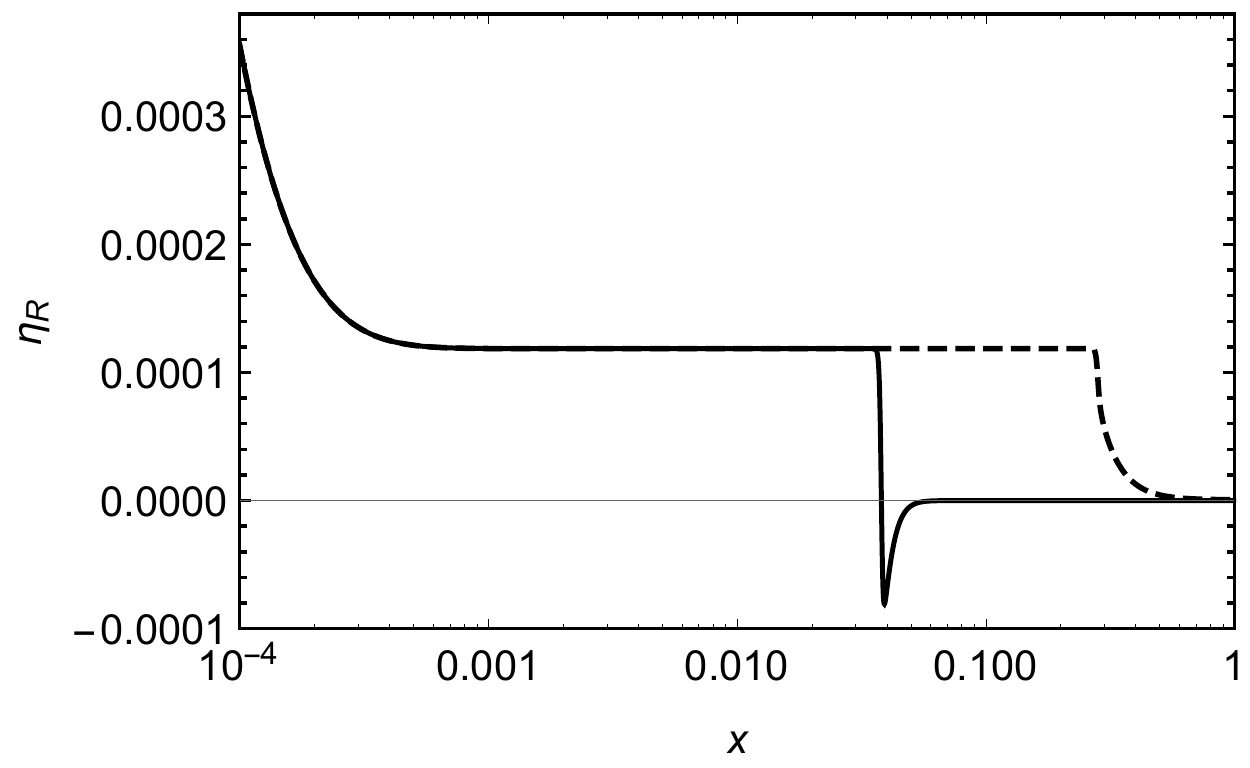}
  \hspace{2mm}
  \includegraphics[width=65mm]{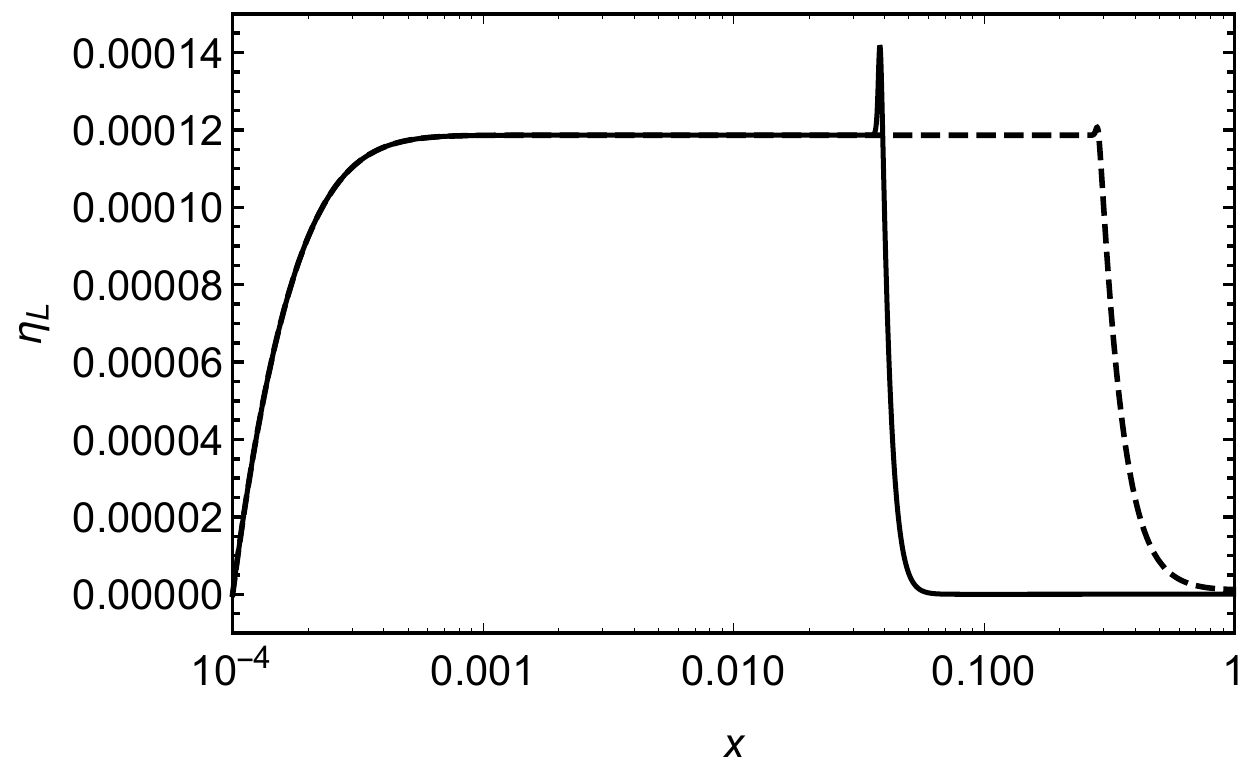}
%\caption{ $E_0=0.001$ and $k=0.5$(Left) and $k=1.5$(Right) }\label{1Jt}
%\end{center}
%\end{figure}
%\begin{figure}
%\begin{center}
  \includegraphics[width=65mm]{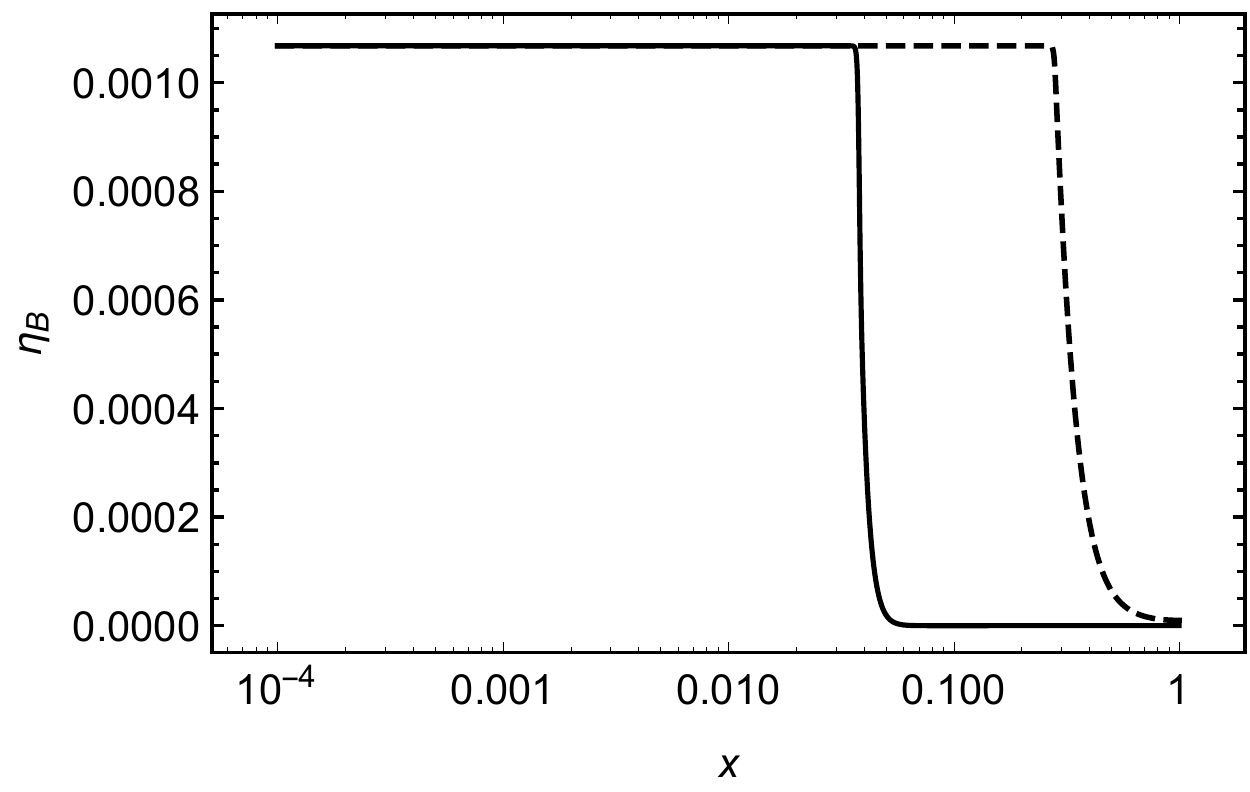}
  \hspace{2mm}
  \includegraphics[width=65mm]{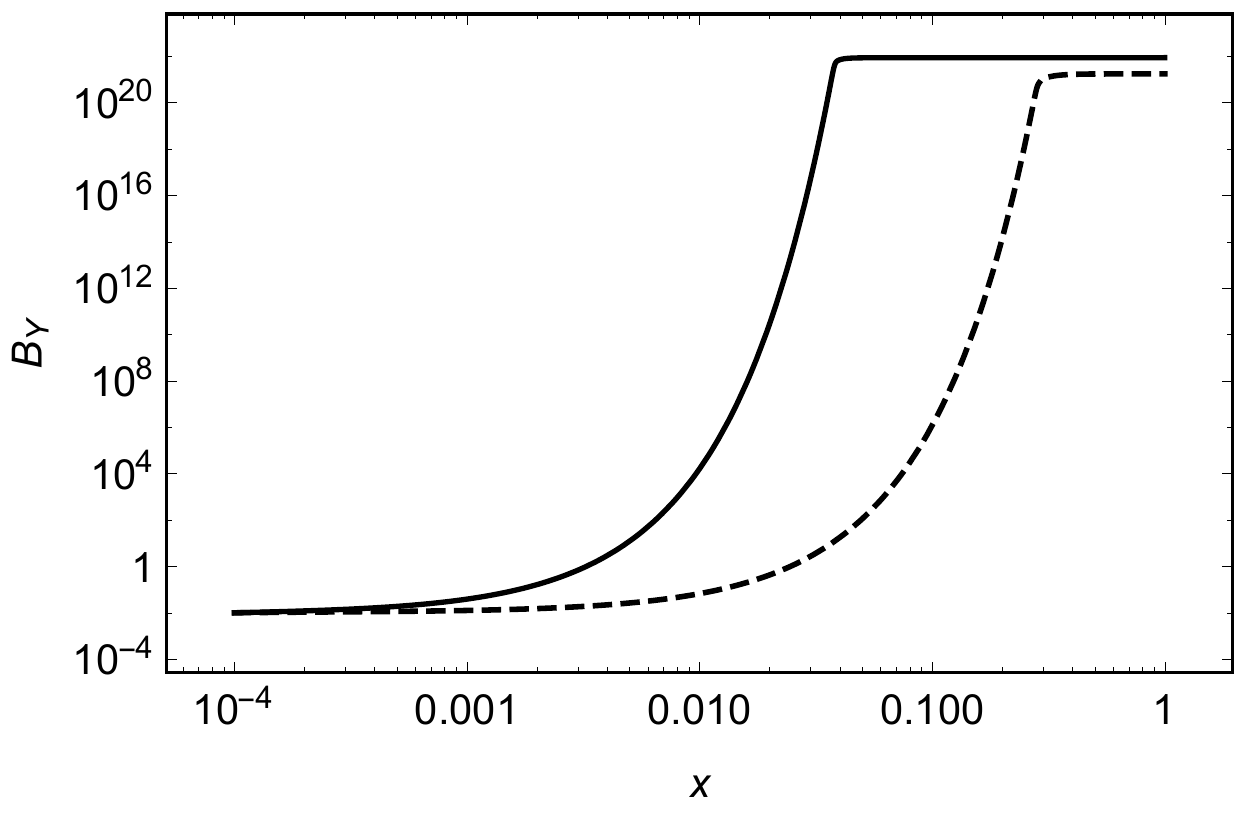}
  %\includegraphics[width=65mm]{logetaBplot2.pdf}
  %\begin{center}
  %\includegraphics[width=65mm]{Byplot2.pdf}
  %\end{center}
\caption{The time plots of the first-generation leptonic asymmetries $\eta_R=\eta_{e_R}$ and $\eta_L=\eta_{e_L}=\eta_{\nu_e^L}$, baryonic asymmetry $\eta_B$, and the hypermagnetic field amplitude $B_Y$, for $k_0=10^{-7} T_{EW}$ with initial conditions $B_Y^{(0)}=10^{-2}G$, and $y_R^{(0)} = 10^{3}$ and $\eta_B^{(0)}$ respecting the condition $\eta_B^{(0)}/3-\eta_{L_e}^{(0)}=0$ for two different cases.\\
 Case 1 (dashed lines): \ \ \  $\alpha_Y^{(0)}=\frac{g'^2}{8\pi^2\sigma}(2\mu_{e_R} - \mu_{e_L})$.\\ Case 2 (solid lines): \ \ \ \ \ \  $\alpha_Y=\frac{g'^2}{8\pi^2\sigma}(2\mu_{e_R} - \mu_{e_L} + \frac{3}{4}\mu_B)$.\\ 
%The starting point is at $x_0=\frac{t_0}{t_{EW}}=(\frac{T_{EW}}{T_0})^2=10^{-4}$ and the final point is at the onset of the electroweak phase transition  $x_f=\frac{t_f}{t_{EW}}=(\frac{T_{EW}}{T_f})^2=1$. Case 1 is obtained from our previous work \cite{shiva} and is reproduced here for comparison purposes. 
 The maximum relative error for these plots is of the order of $10^{-15}$.
%top(left) :\ \ \ \ \ \ \ \ Normalized chemical potential of right-handed electrons, \\
%top(right) :\ \ \ \ \ \ Normalized chemical potential of left-handed electrons, \\ 
%bottom(left) : \ \ Baryon asymmetry, \\
%bottom(right) : Hypermagnetic field amplitude. 
}\label{RGB2}
\end{figure}

%First, the evolution equations are solved numerically by assuming zero initial matter asymmetries but an initial amplitude of the hypermagnetic field $B_Y^{(0)}=10^{21}$G for two different cases, namely $\alpha_Y=\frac{g'^2}{8\pi^2\sigma}(2\mu_{e_R} - \mu_{e_L})$ and $\alpha_Y=\frac{g'^2}{8\pi^2\sigma}(2\mu_{e_R} - \mu_{e_L} + \frac{3}{4}\mu_B)$. The results are presented as time plots in Figure \ref{RGB}. As can be seen, in both cases, matter asymmetry generation occurs in the presence of hypermagnetic fields; however, the final values of the asymmetries at the onset of EWPT for the second case are almost $7$ times smaller than those of the first case. Moreover, the hypermagnetic field amplitude behaves nearly the same with a little more increase in its final value for the second case.

In the next step, we examine the possibility to grow a very weak seed of the hypermagnetic field, e.g. $B_Y^{(0)}=10^{-2}$G, by initial 
baryon and right-handed electron asymmetries which respect the constraint $\eta_B^{(0)}/3-\eta_{L_e}^{(0)}=0$. We solve the evolution equations with $y_R^{(0)} = 10^{3}$ for two different cases, i.e., $\alpha_Y^{(0)}=\frac{g'^2}{8\pi^2\sigma}(2\mu_{e_R} - \mu_{e_L})$ and $\alpha_Y=\frac{g'^2}{8\pi^2\sigma}(2\mu_{e_R} - \mu_{e_L} + \frac{3}{4}\mu_B)$, and display the time plots in Fig. \ref{RGB2}. As can be seen, in both cases, the hypermagnetic field becomes strong in the presence of the initial matter asymmetries. Although the final amplitude of the hypermagnetic field for the second case is about 5 times larger than that of the first case, the final baryon asymmetry is about 40 times smaller as compared to the first case.   
%We detect some differences between these graphs and those of Figure \ref{three-timeplot}. First, both $\eta_R$ and $\eta_L$ finally merge to a much smaller value near zero, and $\eta_B$ drops from its initial amount to a small positive value and never becomes negative. Second, 
Moreover, the anomalous processes which reduce the asymmetries and amplify the hypermagnetic field, start up much sooner in the second case, i.e. near the point $x\sim 0.04$. 
%Moreover, the amplitude of the hypermagnetic field reaches 5 times larger value of about $10^{22}$G, while 

\begin{figure} 
  \includegraphics[width=65mm]{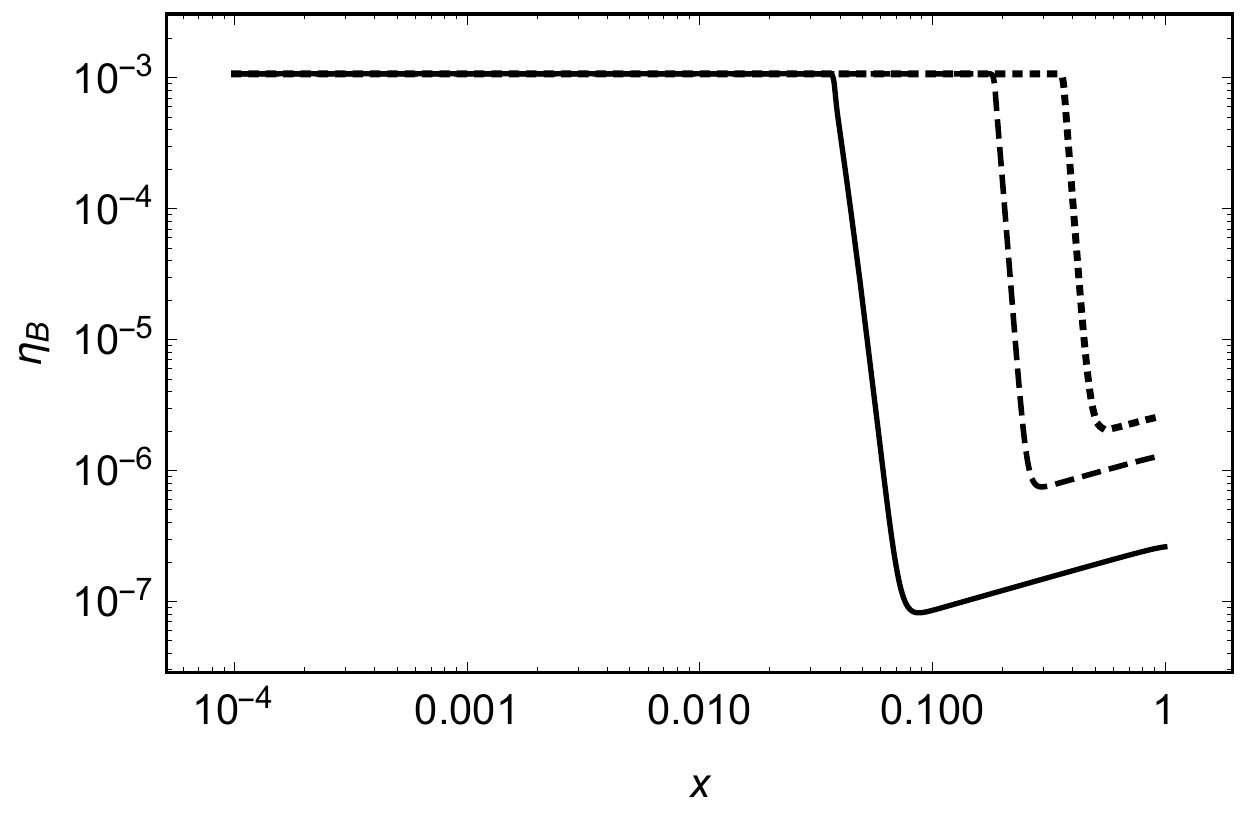}
  \hspace{2mm}
  \includegraphics[width=65mm]{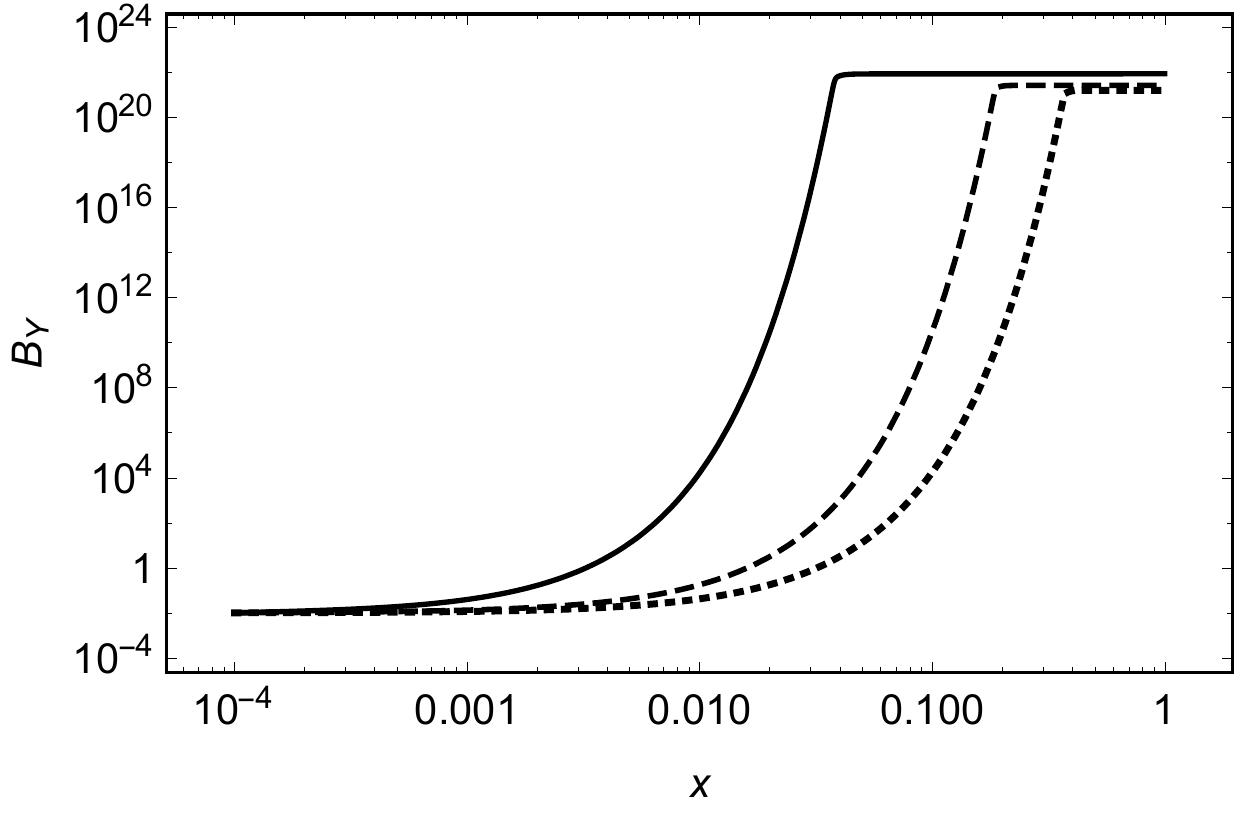}
  \includegraphics[width=65mm]{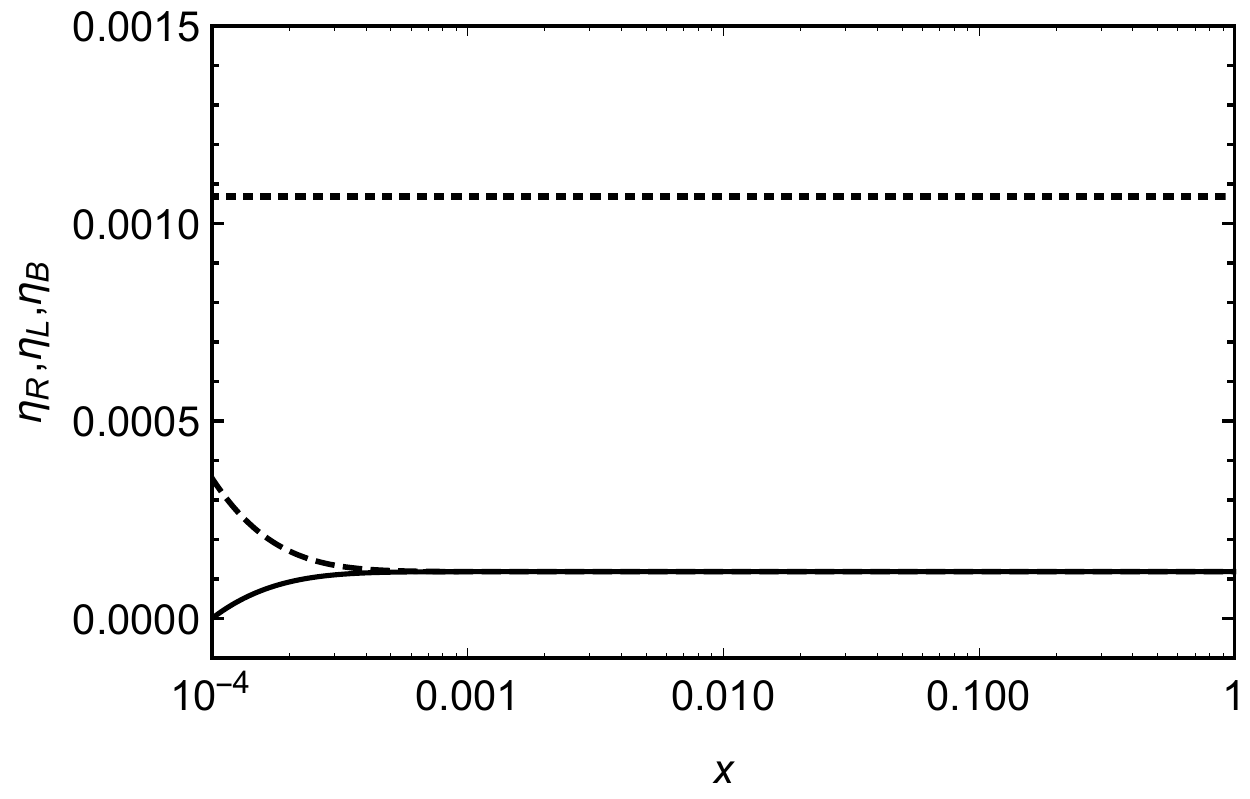}
  \hspace{2mm}
  \includegraphics[width=65mm]{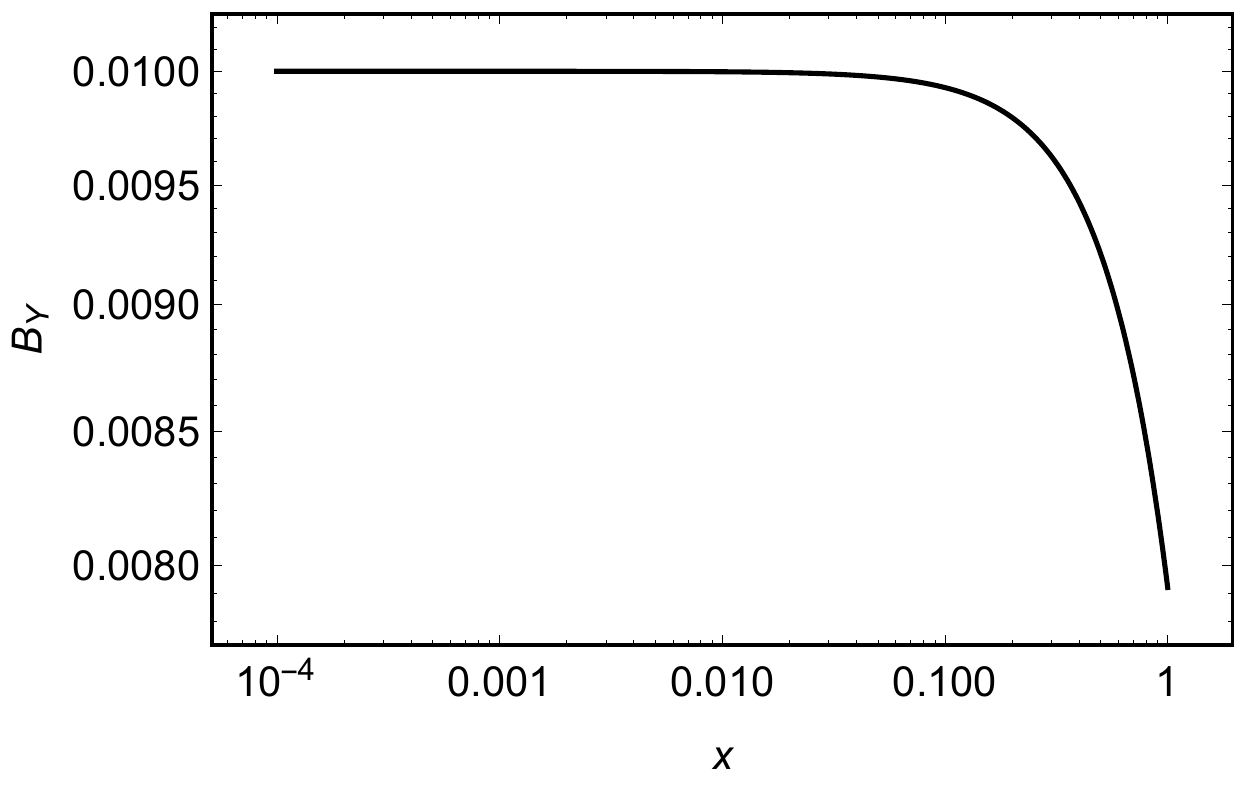}
\caption{Top two plots: The time plots of baryonic asymmetry $\eta_B$ and the hypermagnetic field amplitude $B_Y$ for $k_0=10^{-7} T_{EW}$ with initial conditions $B_Y^{(0)}=10^{-2}G$, and $y_R^{(0)} = 10^{3}$ and $\eta_B^{(0)}$ respecting the conservation law $\eta_B^{(0)}/3-\eta_{L_e}^{(0)}=0$ for three different values of c, namely $c=1$ (solid lines), $c=0.2$ (dashed lines), and $c=0.1$ (dotted lines). The lower two plots are for the case $c=0$ and show the time plots of the first-generation leptonic asymmetries, $\eta_R=\eta_{e_R}$ (dashed line) and $\eta_L=\eta_{e_L}=\eta_{\nu_e^L}$ (solid line), baryonic asymmetry $\eta_B$ (dotted line), and the hypermagnetic field amplitude $B_Y$.  
% Case 1 (solid lines): \ \ \ \ \ \ \ \  $c=1$\\ Case 2 (dashed lines): \ \ \ \ \ $c=0.2$\\Case 3 (dotted lines): \ \ \ \ \ $c=0.1$\\ 
The maximum relative error for these plots is of the order of $10^{-15}$.
}\label{c2}
\end{figure}

Let us again investigate the significance of the U$_\textrm{Y}$(1) Chern-Simons term, whose coefficient is given by Eq.\ (\ref{c'_E2}), via reducing its effect by multiplying it with the adjustable parameter $c\leq1$. We have solved the dynamical equations with the aforementioned initial conditions for three different values of $c:\{0.2,0.1,0\}$ and the resulting time plots, along with the case $c=1$ already obtained, are presented in Fig. \ref{c2}. Again, it can be seen that as the value of $c$ becomes smaller, the final matter asymmetries increase, but the final hypermagnetic field amplitude decreases. More importantly, the case $c=0$ shows that even large matter asymmetries are not able to strengthen the hypermagnetic field in the absence of the U$_\textrm{Y}$(1) Chern-Simons term. 
%Therefore, we obtain the previous result that, when the U$_\textrm{Y}$(1) Chern-Simons term is taken into account, matter asymmetries decrease but the hypermagnetic field gets strong.

Finally, we solve the dynamical equations with $B_Y^{(0)}=10^{-2}$G, $y_R^{(0)}$ in the range $10^{-2}<y_R^{(0)}<10^{3}$ and initial baryon asymmetry fulfilling the condition $\eta_B^{(0)}/3-\eta_{L_e}^{(0)}=0$, and obtain the final values at $T=T_{EW}$. Again we do not display the results for space limitation, and suffice it to point out the salient features of this investigation. 
%This investigation is analogous to the one done in our previous work  and the results are similar. 
%For $y_R^{(0)}$ increasing in the range $10^{-2}<y_R^{(0)}<10^{1.52}$, the final matter asymmetries rise. For $10^{1.52}<y_R^{(0)}<10^{1.74}$, the final baryon asymmetry $\eta_B$ decreases, the final value of $\eta_L$ shows an increase and a decrease, but that of $\eta_R$ decreases then increases, it also obtains negative values. For $10^{1.74}<y_R^{(0)}<10^{3}$, the final values of all matter asymmetries finally merge to much smaller values. Moreover, 
We find that for $10^{-2}<y_R^{(0)}<10^{1.52}$, the final value of the hypermagnetic field amplitude $B_Y$ grows until it becomes as large as $10^{20}$G, then increases with a much smaller slope for $10^{1.52}<y_R^{(0)}<10^{3}$. Moreover, the matter asymmetries stay very close to zero except for $10^{1}<y_R^{(0)}<10^{2}$, where $\eta_B$ and $\eta_L$ attain a maximum and $\eta_R$ attains a negative minimum value close to $y_R^{(0)}\approx 10^{1.55}$.
The behavior described above is somehow similar to the behavior observed in the fifth investigation of our previous work except that, there was no negative value for the final value of $\eta_R$, and the matter asymmetries reach their extremum values around $y_R^{(0)}\approx 10^{2.55}$.
%the critical range of $y_R^{(0)}$ was $10^{2.4}<y_R^{(0)}<10^{2.7}$ instead of $10^{1.52}<y_R^{(0)}<10^{1.74}$. 
Two interesting points can be emphasized about the results: The first one is that at $y_R^{(0)}=10^{1.52}$ strong hypermagnetic field and large amounts of matter asymmetries are obtained at $T=T_{EW}$. Another one is that at $y_R^{(0)}=10^{1.56}$ the final amount of \lq{$\eta_R-\eta_L$}\rq\ becomes maximum. This chiral asymmetry is important for the evolution of Maxwellian magnetic fields in the broken phase \cite{Boyarsky}.

We also repeat the above investigation in the absence of the Abelian Chern-Simons term by choosing $c=0$. We find that the behavior is totally different and none of the interesting features of the previous case can be seen. Indeed, there is no amplification of the hypermagnetic fields. Moreover, the final baryonic asymmetry is the same as its initial value and the final right-handed and left-handed lepton asymmetries are equal, with their sum being equal to $\eta_R^{(0)}$.

\section{Summary and Discussion}\label{Summary and Discussion}
In this paper, we have studied the effect of the U$_\textrm{Y}$(1) Chern-Simons term, and its baryonic contribution, on the evolution of the matter asymmetries and the hypermagnetic field, within the context of a simple model and in the temperature range $100\textrm{GeV}<T<10\textrm{TeV}$. To do the latter, we have compared the results when the coefficient of the U$_\textrm{Y}$(1) Chern-Simons term, i.e. $c'_E$, includes only the usual first generation leptonic contribution, with the results when the baryonic contribution is also included. To study the first part, i.e. the importance of the U$_\textrm{Y}$(1) Chern-Simons term in general, we have studied the effect of multiplying $c'_E$, which now includes the baryonic contribution, by an attenuating parameter $0 \leq c<1$. The baryonic contribution added is subject to the condition $\eta_B^{(0)}/3 - \eta_{L_e}^{(0)} = 0$. One of the effects of this condition is to increase the initial magnitude of $c'_E$. Comparison of the results for the matter asymmetries and hypermagnetic fields with and without the inclusion of the baryonic contribution shows that the results are qualitatively similar. The differences, along with the effect of attenuating the amplitude of the U$_\textrm{Y}$(1) Chern-Simons term to the point of eliminating it altogether, are described below.

We first discuss the generation of matter asymmetries by an initial hypermagnetic field. Our study has shown that an initial non-zero hypermagnetic field can grow matter asymmetries from zero initial value. However, the growth which is initially quadratic with respect to $B_Y^{(0)}$, saturates for values larger than a critical value denoted by $B_{Y,C}^{(0)}$. Therefore the larger $B_{Y,C}^{(0)}$, the larger the final saturated values of the matter asymmetries. The values of $B_{Y,C}^{(0)}$, for the cases with and without the baryonic contribution are approximately $10^{20.5}$G and $10^{21}$G, respectively, leading to about seven times smaller final matter asymmetries in the first case. This comparison also indicates that $B_{Y,C}^{(0)}$ increases with attenuating $c'_E$, a conclusion which is confirmed with the use of attenuating parameter. In this regard, the interesting point is that when the Chern-Simons term is eliminated altogether by setting $c=0$, the growth of the matter asymmetries continue to be quadratic with respect to $B_Y^{(0)}$ without any saturation, as though $B_{Y,C}^{(0)}$ has moved to infinity. On the other hand, the change in final value of the hypermagnetic field, denoted by $B_Y (T_{EW} ) $, is very small in either case. For the case shown in Fig.\ \ref{RGB}, when the baryonic contribution is added it increases by 1\%, as compared to 0.2\% when it is not. Both of these cases are indications of a mild resonance. Moreover, as the attenuating parameter $c$ decreases, $B_Y (T_{EW} ) $ decreases as well, becoming equal to its initial value for $c\approx0.1$, and decreasing by 20\% when $c=0$. 

Next, we discuss the generation of hypermagnetic field by an initial matter asymmetry. As mentioned before, the generation of a nonzero $B_Y (T_{EW} ) $ is possible only if its initial value is non-zero. The time plots show that in general one can identify a particular time, denoted by $t_{\textrm{Tr}}$, where the important transitions start. Figure \ref{RGB2} shows that the inclusions of the baryonic contribution leads to a decrease in $t_{\textrm{Tr}}$, i.e. the transitions start at a higher temperature. Moreover, at $t_{\textrm{Tr}}$ the matter asymmetries drop rather sharply, and the growth of the hypermagnetic field, which had been steady heretofore, saturates. For the case displayed, $B_Y (T_{EW} ) $ becomes about five times larger when the baryonic contribution is included, while the final matter asymmetries become about forty times smaller. Figure \ref{c2}, which displays the effects of the attenuation parameter, shows that the features just described are generic consequences of changing the value of $c'_E$. Figure \ref{c2} also show a very interesting case of $c=0$. In this case the matter asymmetries do not change, except for balancing out due to chirality flip processes. More importantly, the minute the hypermagnetic field seed not only does not grow but drops by 20\%. Another interesting outcome of the investigation which includes the range $10^{-2}\leq y_R^{(0)}\leq 10^{3}$ is that, when no attenuation parameter is taken into account, almost all of the matter asymmetries are expended to grow $B_Y (T_{EW} ) $. Here the point $ y_R^{(0)} \approx 10^{1.5}$ stands out around which the rate of growth of $B_Y (T_{EW} ) $ suddenly slows down considerably and the final matter asymmetries attain their extremum values. Surprisingly, the extremum of $\eta_R$ is a negative minimum. Hence a relatively large chiral asymmetry is generated at this point, which is important for the subsequent evolution of the Maxwellian magnetic field in the EWPT and the broken phase. The corresponding point in the absence of baryonic contribution is $ y_R^{(0)} \approx 10^{2.4}$.

We mentioned in Section \ref{Introduction} that the baryon asymmetry of the Universe (BAU) is $\eta_B\sim 10^{-10}$ as extracted from the observational data. Let us also briefly state some features of the observational data about the magnetic fields, and then check the compatibility of our results with these data. 

%The observations of the CMB temperature anisotropy put an upper bound on the strength $B_0$ of the present magnetic fields, $B_0 \lesssim 10^{-9}$G on the CMB scales $\lambda_0\gtrsim 1$Mpc \cite{Ade}. Furthermore, the observations of the gamma rays from blazars not only provide both the lower and upper bounds on the strength $B_0$ but also indicate the existence of the large scale magnetic fields with the scales as large as $\lambda_0\simeq1$Mpc \cite{Ando,Essey,Chen}. The strength $B_0$ of the present intergalactic magnetic fields (IGMFs) reported in \cite{Ando} is $B_0\simeq10^{-15}$G. Two different cases are also investigated in Ref.\ \cite{Essey}. In the first case, where blazars are assumed to produce both gamma rays and cosmic rays, they find $1\times 10^{-17}\textrm{G}<B_0<3\times10^{-14}\textrm{G}$. However, in the second case where the cosmic ray component is excluded, they report that the $10^{-17}$G lower limit remains valid but the upper limit depends on the spectral properties of the source. Ref.\ \cite{Chen} estimates the strength of the IGMFs to be in the range $B_0\simeq 10^{-17}-10^{-15}$G, which is consistent with the above mentioned results of \cite{Ando,Essey}. Moreover, a nonvanishing helicity of the present large scale magnetic fields is also infered with the strength $B_0\simeq5.5\times 10^{-14}$G in Ref.\ \cite{Chen2}. We do not know whether these magnetic fields are entirely primordial, but if they are, then they are the closest measure of the primordial fields.

The observations of the CMB temperature anisotropy put an upper bound on the strength $B_0$ of the present magnetic fields, $B_0 \lesssim 10^{-9}$G on the CMB scales $\lambda_0\gtrsim 1$Mpc \cite{Ade}. Furthermore, the observations of the gamma rays from blazars not only provide both lower and upper bounds on the strength $B_0$, but also indicate the existence of the large scale magnetic fields with the scales as large as $\lambda_0\simeq1$Mpc \cite{Ando,Essey,Chen}. The strength $B_0$ of the present intergalactic magnetic fields (IGMFs) reported in \cite{Ando} is $B_0\simeq10^{-15}$G. Two different cases are also investigated in Ref.\ \cite{Essey}. In the first case, where blazars are assumed to produce both gamma rays and cosmic rays, they find $1\times 10^{-17}\textrm{G}<B_0<3\times10^{-14}\textrm{G}$. However, in the second case where the cosmic ray component is excluded, they report that the $10^{-17}$G lower limit remains valid but the upper limit depends on the spectral properties of the source. Reference \cite{Chen} estimates the strength of the IGMFs to be in the range $B_0\simeq 10^{-17}-10^{-15}$G, which is consistent with the above mentioned results of \cite{Ando,Essey}. Moreover, a nonvanishing helicity of the present large scale magnetic fields is also infered with the strength $B_0\simeq5.5\times 10^{-14}$G in Ref.\ \cite{Chen2}.

Aside from the cosmic expansion which leads to the trivial adiabatic evolution of the cosmic magnetic fields, several other effects such as the viscous diffusion, the inverse cascade, the Abelian anomalous effects, etc, affect their evolution as well. In the trivial case, the strength $B(t)$ and the scale $\lambda(t)$ are proportional to $a^{-2}(t)$ and $a(t)$, respectively, where $a(t)$ is the FRW scale factor. However, in the inverse cascade mechanism, %which needs large amounts of magnetic helicity to operate correctly, 
$\lambda(t)$ grows faster than $a(t)$ due to the turbulence in the plasma \cite{Fujita}. In this case, the magnetic helicity is approximately conserved but the energy is transferred from small scales to large scales \cite{Kahniashvili}, and the spectrum develops with a characteristic scaling law \cite{Campanelli}. After recombination, the plasma becomes neutral and the magnetic fields evolve trivially. One can use the scaling relation to express the spectrum of the primordial magnetic fields in terms of $\lambda_0$ and $B_0$ as (see Ref.\ \cite{Fujita} and Appendix C of Ref.\ \cite{Kamada})
\be\begin{split}\label{lambda-B}
B(T)\simeq(1\times10^{20}\textrm{G})(\frac{T}{100\textrm{GeV}})^{7/3}(\frac{B_0}{10^{-14}\textrm{G}})g_B(T),\cr
\lambda(T)\simeq(2\times10^{-29}\textrm{Mpc})(\frac{T}{100\textrm{GeV}})^{-5/3}(\frac{\lambda_0}{1\textrm{pc}})g_{\lambda}(T) 
\end{split}\ee
where $g_B(T)$ and $g_{\lambda}(T)$ are O(1) factors. The following linear relation can also be obtained for the magnetic fields that have experienced the inverse cascade process \cite{Banerjee,Durrer} 
\be\label{lambda0B0}
\frac{\lambda_0}{1\textrm{pc}}\simeq a\frac{B_0}{10^{-14}\textrm{G}},
\ee 
where the range of $a$ is inferred to be from O(0.1) to O(1) \cite{Kamada}. Let us now use these inverse cascade results to see whether our results are compatible with the observations.

The inverse cascade mechanism that we want to invoke in the broken phase, needs magnetic helicity in order to operate efficiently. So, let us first investigate whether our helical hypermagnetic field leads to a helical Maxwellian magnetic field after the electroweak phase transition, via calculating the magnetic helicity before and after the symmetry breaking. In the symmetric phase, the hypermagnetic helicity is defined as 
$\overline{\textbf{Y}.\textbf{B}_\textbf{Y}}$, where the overline represents the volume average. We calculate this quantity for our simple wave configuration of the hypermagnetic field and obtain $\overline{\textbf{Y}.\textbf{B}_\textbf{Y}}=\textbf{Y}.\textbf{B}_\textbf{Y}=k_0y^2(t)=B_Y^2(t)/k_0$.
During Standard Model electroweak symmetry breaking, 3 out of 4 gauge fields of SU(2)$_L\times$ U(1) acquire mass, i.e. $W^{\pm}\ \textrm{and}\ Z$, while one combination, i.e. photon, remains massless.
%the weak gauge fields get mass and the hypermagnetic field is transformed into an electromagnetic one. 
A thorough study of this evolution in the plasma of the early Universe is beyond the scope of this work. Therefore, we choose the following simple model presented in Sec.\ 2 of Ref.\ \cite{Kamada} which assumes that the system passes abruptly from the symmetric phase to the broken phase (in a way similar to that of Ref.\ \cite{Pavlovic}). Then, we can estimate the strength $B$ and the magnetic helicity $\overline{\textbf{A}.\textbf{B}}$ of the magnetic field after the symmetry breaking. Let us recall the relations:
\be\begin{split}\label{ZA}
Z_{\mu}= c_W W_{\mu}^3-s_W Y_{\mu},\cr
A_{\mu}= s_W W_{\mu}^3+c_W Y_{\mu},
\end{split}\ee
where $s_W$ and $c_W$ are the sine and cosine of the weak mixing angle $\theta_W$, and $s_W^2=0.23$. It can be seen that the hypermagnetic field $\textbf{B}_\textbf{Y}$ has components in both $\textbf{B}_\textbf{Z}$ and $\textbf{B}_\textbf{A}$. As the Higgs condensate grows at the EWPT, the W and Z fields get mass and decay. Following the simple model presented in Ref.\ \cite{Kamada}, we assume that the Z component of $\textbf{B}_\textbf{Y}$ decays rapidly at the EWPT. Therefore, the $\textbf{B}_\textbf{Z}$ component of $\textbf{B}_\textbf{Y}$ vanishes and the electromagnetic component $\textbf{B}_\textbf{A}$ remains. Moreover, the thermal expectation value $\langle W_{\mu}^a \rangle = 0$, since in the symmetric phase the non-Abelian gauge fields $W_{\mu}^a (x)$ acquire mass from their self-interactions in the plasma \cite{Gross} and are screened. Then, we obtain the electromagnetic component in the form, $\textbf{E}= c_W \textbf{E}_\textbf{Y}$ and $\textbf{B}= c_W \textbf{B}_\textbf{Y}$. This means that the strength decreases by about $10\%$ ($B\simeq0.88B_Y$) and the magnetic helicty decreases around $20\%$ ($\overline{\textbf{A}.\textbf{B}} \simeq 0.77\  \overline{\textbf{Y}.\textbf{B}_\textbf{Y}}$). Although the helicity is decreased, the Maxwellian magnetic fields of the broken phase are still helical. Hearafter, we consider the simplifying assumption of neglecting the decrease in the magnitudes of these quantities, since it does not significantly affect our order of magnitude estimates of the strength $B_0$ and the scale $\lambda_0$ of present magnetic fields.

Using the relation $\lambda=k_0^{-1}$, the scale of the hypermagnetic field used in our investigations is estimated as $\lambda{(T_{EW}\simeq 100\textrm{GeV})}=(10^{-7}T_{EW})^{-1}=6.45\times10^{-28}\mbox{pc}$.
%Let us estimate roughly the scale of the hypermagnetic field used in our investigations via the relation $\lambda=k_0^{-1}$:
%\be
%\lambda_{(T=100\textrm{GeV})}=(10^{-7}T_{EW})^{-1}=6.45\times10^{-28}\mbox{pc},
%\ee
%which is a very small scale.
Let us first assume that the magnetic fields evolve trivially from EWPT till present ($T_0\simeq 2\textrm{K}\simeq17.2\times10^{-14}\mbox{GeV}$). Then, using the mentioned relation $\lambda(t) \propto a(t) \propto T^{-1}$, the present scale of the magnetic fields is obtained as
%Assuming that the time evolution of magnetic field from $T=100$GeV to $T=2\mbox{K}=17.2\times10^{-14}$GeV is trivial, we can compute the resultant scale of the magnetic field at present time via the mentioned relation $\lambda(t) \propto a(t) \propto T^{-1}$:
\be\begin{split}
\lambda(T_0)=\lambda(T_{EW})\left(\frac{\mbox{100GeV}}{17.2\times10^{-14}\mbox{GeV}}\right)\simeq3.75\times10^{-13}\mbox{pc},\cr
\end{split}\ee
which is much lower than the acceptable scales of present magnetic fields. When we decrease the wave number $k_0$ to $10^{-3}k_{max}$, the saturated value of the baryonic asymmetry mentioned in Subsection \ref{Matter Asymmetry Generation by Hypermagnetic Fields} becomes $\eta_B\simeq10^{-10}$. Indeed, no wave number lower than this one can give the BAU in our model. The scale $\lambda$ corresponding to this $k_0$ is $\lambda(T_{EW})\simeq6.45\times10^{-25}\mbox{pc}$ leading to $\lambda(T_0)\simeq3.75\times10^{-10}\mbox{pc}$, which is still far from the current scales of magnetic fields. These calculations show that for obtaining the present large scale magnetic fields, it is necessary to rely on an inverse cascade process which starts after the EWPT.

Let us assume that the inverse cascade process is the only nontrivial process which starts immediately after the EWPT. Then, using Eqs.\ (\ref{lambda-B}), and Eq.\ (\ref{lambda0B0}) with $a\simeq 0.1$, we can roughly estimate $\lambda_0$ and $B_0$ for $\lambda(T_{EW})\simeq6.45\times10^{-25}\mbox{pc}$ ($k_0=10^{-3}k_{max}$) and $B(T_{EW})\simeq 3.225\times 10^{19}\mbox{G}$ to obtain 
\be\begin{split}
B_0\simeq 3.225\times 10^{-15}\mbox{G},\ \ \ \textrm{and} \ \ \ \lambda_0\simeq 3.225\times 10^{-2}\mbox{pc}.
\end{split}\ee

%Let us assume the minimum $\lambda_0\simeq1$pc which corresponds to $B_0\simeq10^{-14}$G, and assume that the inverse cascade process starts immediately after the EWPT. Then, we can roughly estimate $\lambda$ and $B$ at $T=100$GeV by using the equations (2.4) and (2.5) from Ref.\ \cite{Fujita} to obtain
%\be\begin{split}
%B_{(T=100\textrm{GeV})}\simeq9.3\times10^{19}\mbox{G},\ \ \ \ \ \ \ \lambda_{(T=100\textrm{GeV})}\simeq 2.4\times 10^{-23}\mbox{pc}
%\end{split}\ee

%It can be seen that the above estimated strength $B_0$ is consistent with the ones observed; however, the above estimated scale $\lambda_0$ is much smaller than the assumed scale of present magnetic fields which is about $\sim1$Mpc \cite{Ando,Essey,Chen}. Therefore, the above strength of the magnetic fields along with $\eta_B\sim10^{-10}$ already used are within the acceptable range of present day data. However, the maximum scale of present magnetic fields which we can obtain with our simple model and the aforementioned assumptions is much lower than what is usually assumed for gamma ray propagation.  

It can be seen that the above value of $B_0$, along with $\eta_B \sim 10^{-10}$ already used, are within the acceptable range of present day data. However, the value of $\lambda_0$ is much smaller than the scale usually assumed for gamma ray propagation which is about $\sim1$Mpc \cite{Ando,Essey,Chen}.

The above results are obtained using a single-mode wave configuration of the hypermagnetic field which is maximally helical, since its helicity density $h_Y=\textbf{Y}.\textbf{B}_\textbf{Y}=k_0y^2(t)$ is related to its energy density $\rho_Y=\textbf{B}_\textbf{Y}.\textbf{B}_\textbf{Y}/2=k_0^2y^2(t)/2$ via the relation $k_0h_Y=2\rho_Y$ (or equivalently, $\rho_k=\frac{k}{2}h_k$ in Fourier-space). The use of this field configuration seems to be an oversimplification; however, as we shall argue below, it is adequate for our purposes. As mentioned in Sec.\ 1, a helical magnetic field may have been generated during the inflation (see also Ref.\ \cite{Giovannini2}).
%we have the assumption of helical magnetic field generation during the inflation [Giovannini]. 
Even if the generated field is partially helical, it would become maximally helical through an inverse cascade mechanism after the inflation \cite{Saveliev}. Nevertheless, let us predict the consequences of choosing a more complicated initial field configuration; namely, a superposition of the fields with different values of $k_0$.

To accomplish this task, we first study an analogous case in the broken phase, which investigates the evolution of the magnetic fields, taking into account the chiral anomaly \cite{Boyarsky}. It has been shown that for a continuous spectrum magnetic field, a very important effect emerges; that is, the initial spectrum reddens with time, while the total helicity remains (nearly) conserved, similar to the well studied turbulence-driven inverse cascade phenomenon for the helical magnetic fields. However, in this case, the magnetic energy and helicity transfer from shorter to longer scales occur not because of the turbulence but due to the chiral anomaly. In continuation, the authors of \cite{Boyarsky} have analyzed a special helical single-mode solution of the system of chiral MHD equations (exactly like our simple wave configuration), and have shown that their qualitative conclusions reached in \cite{Boyarsky} remain valid \cite{Boyarsky2}. In particular, they have shown an important property of the helical single-mode solutions in the presence of a homogeneous axial chemical potential, which is the “inverse cascade” phenomenon, i.e., the transfer of energy and magnetic helicity from short to large scales.

Similar works have also been done in the symmetric phase which show the same effect \cite{Smirnov,Tashiro}. Indeed, the evolution equations of the hypermagnetic fields and the fermionic chemical potentials, taking into account the Abelian anomalous effects in the symmetric phase, are similar to those of the magnetic fields and the axial chemical potential ($\Delta\mu=\mu_L-\mu_R$) considering the chiral anomalous effects in the broken phase. Therefore, it seems that for the superposition of the fields with different values of $k_0$ as an initial configuration, a fast decay of one helicity mode and an exponential growth of its adjacent lower helicity mode occurs, while the total helicity remains constant. This also leads to the total magnetic energy dissipation, since $\rho_k=\frac{k}{2}h_k$ for helical fields. Finally, the helicity concentrates around the longest mode which can be chosen to be the $k_0$ studied in this paper.
%$k_n$, and $\eta_B$ becomes $\eta_B \propto k_n$. Therefore, it seems possible to obtain the previous results starting with a superposition of the helical fields. In particular, our desired helical hypermagnetic field with the single mode $k_0=10^{-3}k_{max}$ and the amplitude $B_Y(T_{EW})=3.225\times10^{19}$G (or $\rho_Y(T_{EW})=B_Y^2(T_{EW})/2$) which leads to the baryonic asymmetry $\eta_B\sim10^{-10}$ at the EWPT can be obtained provided that $k_0$ is chosen as the longest mode and the total initial energy density is much higher than its value at the EWPT. 
Therefore, the study of the single-mode can reveal the important features of the system and imply the behavior of the system in the presence of more complicated configurations of the hypermagnetic field.\\

\noindent Acknowledgements: We would like to thank the research office of the Shahid Beheshti University for research facilities.

%\newpage

\end{document}